(18k words inclusive; **Figure 12**; 185 refs.Table 5; updated 5/25/05)

# Mechanical strength of atomic chains, surface skins, and nanograins

Chang Q. Sun[*]

*School of Electrical and Electronic Engineering, Nanyang Technological University, Singapore 639798, Singapore*

This report deals with the correlation between the mechanical strength and thermal stability of systems extending from monatomic chains to surface skins and solids over the whole range of sizes with emphasis on the significance of atomic coordination imperfection. Derived solutions show that a competition between the bond order loss and the associated bond strength gain of the lower coordinated atoms dictate the thermo-mechanics of the low dimensional systems. Bond order loss lowers the atomic cohesive energy that determines the temperature of melting ($T_m$), or the activation energy for atomic dislocation, whereas bond strength gain enhances the energy density, or mechanical strength, in the surface skin. Therefore, the surface is harder at $T << T_m$ whereas the surface becomes softer when the T approaches the surface $T_m$ that is lower than the bulk due to bond order loss. Hence, the strained nanostructures are usually stiffer at low T whereas the harder skins melt easier. Quantitative information has been obtained about the bonding identities in metallic monatomic chains and carbon nanotubes. Solutions also enable us to reproduce the inverse Hall-Petch relationship with clarification of factors dominating the transition from hardening to softening in the nanometer regime.

Keywords: nanostructures; surfaces; monatomic chain; mechanical strength; thermal stability; extensibility; Hall-Petch relationship.

---

[*] Fax: 65 6792 0415; E-mail: ecqsun@ntu.edu.sg; URL: www.ntu.edu.sg/home/ecqsun/

I   Introduction

1.1   Scope

This report starts in Section 1 with an overview on the unusual behavior of systems ranging from surfaces, monatomic chains (MCs), and nanobeams (tubes and rods) to solids in nanometer and micrometer regimes in mechanical strength and thermal stability with discussion on possible mechanisms and challenging issues. Section 2 describes the principle of bond order-length-strength (BOLS) correlation and its derivatives on the analytical solutions to the corresponding issues. Section 3-5 compares theoretical predictions with experimental measurements on the low-dimensional systems. Section 6 discusses further with evidence for the approach and the impact of atomic coordination imperfection in dominating mechanical strength and thermal stability of low dimensional systems. Section 7 summarizes the achievement and limitations of the solutions with brief discussion on perspectives.

1.2   Overview
1.2.1   Surfaces

Nanosolids (or termed as nanoclusters, nanocrystallites, etc., have attracted tremendous interest, since they show intriguing properties from a basic scientific viewpoint, as well as great potential in upcoming technological applications such as nanoelectronic devices.[1] The unusual behavior of a flat surface and a nanosolid with curved surface in mechanical strength such as hardness (P, also termed as yield strength or flow stress), Young's modulus (Y), and compressibility ($\beta$, also ductility or extensibility) and thermal stability (solid-semisolid and semisolid-liquid transitions) has been a subject of intensive study during past decades. The enhancement of mechanical strength and Young's modulus of a surface and a nanosolid and the suppression of thermal stability have been widely reported with debating mechanisms. For example, the hardness of nanocrystalline and microcrystalline nickel, at room temperature, varies not only with the geometrical shapes (conical, berkovich, and cube-corner) of the indenter tips but also with the strain rate in measurement though the former is more significant. The maximal hardness that is peaked at a surface depth of ~5 nm varies in a range of 6 and 20 GPa with the tip shapes but the peak position (5 nm depth) changes with neither the shapes of tips nor the strain rate.[2] The values of both the Y and the P for nitrogen-doped amorphous carbon films at the ambient temperatures also show maximal hardness near the surface. The maximal hardness is 3 ~ 4 times higher than the bulk values whereas the peak positions in the hardness-depth profiles remain unchanged when the nitrogen contents or film thickness is varied.[3] Surfaces of Ag, Ni, Cu, and Al thin films are 4 ± 0.5 times harder than the bulk and the hardness of $\alpha_2$-TiAl and $\gamma$-TiAl surfaces is ~2 times the corresponding bulk values.[4] The hardness of Ti, Zr, and Hf carbide films on silicon substrate increases from the bulk value of 18 to 45 GPa when the film thickness is decreased from 9000 to 300 nm.[5] The Young's modulus of nanograined steel was determined to increase from 218 to 270 GPa associated with mean lattice contraction from 0.2872 to 0.2864 nm when the grain size is reduced from 700 to 100 nm.[6] By squeezing Si nanospheres of different sizes between a diamond-tipped probe and the sapphire surface, Gerberich et al.[7] determined at room temperature that a defect-free silicon nanosphere with a diameter of 40 nm is ~3 times (50 GPa) harder than bulk silicon (12 GPa). The smaller the sphere, the harder it was: spheres with a diameter of 100 nm had a hardness of around 20 GPa. Nanoporous Au composed of 10 - 60 nm sized polycrystallines under compressive stress demonstrates the ductile deformation behavior. A mean hardness of 145 ± 11 MPa and Young's modulus of 11.1 ± 0.9 GPa, being 10 times higher than the hardness predicted by scaling laws of open-cell foams and the same scales of intrinsic gold.[8] The melting point suppression with solid size is realized in a form of liquid skin nucleation and growth, which has been intensively discussed in the literature.[9,10] It is general that a flat or curved surface is harder than the bulk interior at temperature far below the melting point, while the surface skin melts prior to the solid core, which is intrinsic disregarding the materials, testing methods, or experimental conditions.

Under mechanical loading, the elastic response is known to be different. Experimental studies have shown that nanoplates or free standing films can be stiffer[11] or softer[12] than their bulk counterparts. The understanding of the softening and stiffening phenomena has evolved over the years. Wolf[13] has shown, using linear elasticity analyses and molecular statics simulations, that surface stress leads to decrease of atomic volume that is correlated with stiffening phenomena. Huang et al[14,15] suggested that the stiffening or softening of metal nanoplates is the result of competition between bond loss, and bond saturation or electron redistribution in addition to surface reconstruction, or alignment of bond chains, or oxidation. Therefore, the mechanism of size induced softening or stiffening is underdebate.

1.2.2 Monatomic chains

A metallic monatomic chain (MC) is an ideal prototype of nanowires for extensibility study, as the extension of an MC involves only bond stretching without bond unfolding or atomic gliding dislocations, as do atoms in coarse-grained metallic chunks under external mechanical stimuli.[16] The intriguing phenomena appeared to MCs include quantum conductance, higher chemical reactivity, lower thermal stability, and the unusually high mechanical strength and ductility. The quantum conductance has been well understood as due to the enlarged sublevel separation, or termed as Kubo gap, $\delta_k = 4E_F/3N$, where $E_F$ is the bulk value of Fermi energy and N is the total number of atoms of the solid in the conduction band.[17,18] Knowledge about the equilibrium bond length, bond strength, extensibility, and the thermal and chemical stability of MCs is yet quite preliminary and controversy.

For instances, the stretching limit of Au-Au distance in the gold MC has been measured using transmission electron microscopy (TEM) at room temperature to vary from 0.29 nm,[19] 0.36 nm (± 30%),[20] 0.35 ~ 0.40 nm[17] and even to a single event of 0.48 nm.[21] However, at 4.2 K the breaking limit of the same Au-Au bond is reduced to 0.23 ± 0.04 nm as measured using scanning tunneling microscopy (STM) and to 0.26 ± 0.04 nm as measured using mechanically controllable break junctions,[22] with respect to the bulk value of 0.2878 nm. Sophisticated density functional theory calculations[23] suggest that the pair-wise interatomic potential is applicable and the Au-Au equilibrium distance (without external stimulus) is between 0.232 and 0.262 nm and the cohesive energy per bond changes from –0.51 to –1.59 eV. The similar trends also hold for other metallic MC such as Pt, Cu, and Ag in theory though experimental observation of such MCs formation at room temperature has been not so frequently. Unfortunately, numeral calculations for pure gold MC have yielded a maximum Au-Au distance of 0.31 nm under tension,[24,25,26,27] which is inconsistent with the values measured at the ambient temperatures in ultra-high vacuum. According to calculations,[28] the mechanical strength of the Au-Au bond is about twice that of the bulk value. The discrepancy of the measured Au-Au distances could not be theoretically solved without light impurity atoms such as X (= H, B, C, N, O, and S, and CO) being artificially inserted into the Au-Au chain.[1] The light atom insertion has led to gold separation in the Au-X-Au chain that could finally match various observed Au-Au distances. For instance, the insertion of a carbon atom led to the stretched Au-C-Au distance of 0.39 nm just before breaking; wires containing B, N, and O displayed even larger distances under tension. The Au-H-Au distance, 0.36 nm, matches one of the experimentally measured values, and the anomalously large distance of 0.48 nm matches the separation between gold atoms in an Au-S-Au chain.[29] Besides the density functional theory approaches, Jiang et al[30] correlated the formation tendency of a MC under tensile stress to the ratio between Peierls stress of a bulk crystal having dislocations and the theoretical break shear stress of the MC. They suggest that the metallic elements having the largest Poisson's ratio hold the largest MC forming ability since such metals have the smallest elastic energy storage within the crystals and can thus endure the largest plastic deformation.

1.2.3 Nanotubes and nanorods

Nanobeams such as hollow tubes and solid fibers have many potential applications[31] in many areas. Since the discovery of CNTs,[32] there has been ever-increasing interest in the new form of carbon because of not only the novel structures and the properties but also the potentially important applications such as atomic-force microscope tips,[33] field emitters and radiation sources,[34,35] electronic devices,[36,37,38,39] hydrogen storage,[40] and chemical sensors.[41] SWCNTs have exceptionally high strength and stiffness[42] and high thermal and electrical conductivity,[43,44] making them excellent candidates for aerospace structural materials.[45] Some of these applications have been now realized in products.

The strong bonds between adjacent carbon atoms make individual nanotubes one of the toughest materials known. A rather stiff CNT gel can be made by physically grinding up SWCNTs with ionic liquids. The stiffness results from a physical cross-linking of the tubes with the ionic liquid. The gels showed good thermal and dimensional stability and could be shaped into conductive sheets that also had enhanced mechanical properties. Using a coagulation-based carbon nanotube spinning technique, Dalton et al[46] spun surfactant-dispersed SWCNTs from a rotating bath of aqueous polyvinyl alcohol to produce nanotube gel fibers that they then converted to solid nanotube composite fibers. The resulting 100 m long fibers were 50 μm thick and contained around 60% nanotubes by weight. They had a tensile strength of 1.8 GPa and an energy-to-break value of 570 J/g. The fibers, which are suitable for weaving into electronic cloth, are four times tougher than spider silk and 17 times tougher than the Kevlar fibers used in bulletproof vests. The fibers also have twice the stiffness and strength and 20 times the toughness of the same weight of a steel wire. As artificial muscles, nanotube fibers are some about 100 times the force of natural muscle with the same diameter.[47] The toughness of spider silk that is dominated by hydrogen bond[48] is due to chain extension in amorphous regions between relatively rigid crystalline protein blocks. The polyvinyl alcohol, with a largely amorphous coating on the nanotubes, is suggested to serve a similar function in the carbon nanotube composite fibers.

However, determination of the Young's modulus of the CNTs has been a confusing issue for a long time. It has been reported that the Y values vary over a range of 0.5 - 5.5 TPa subjecting to the presumption of wall thickness. The Y value for the bulk graphite or diamond is 1.02 TPa. In practice, one can only measure the product of Young's modulus and wall thickness, Yt, rather than the individual component, Y or t. Therefore, the measured Y values are not certain. If one assumes the equilibrium interlayer spacing of graphite sheet as $t_1 = 0.34$ nm, to represent the single-wall (bond) thickness, the derived $Y_1$ value is ~1.1 TPa.[49,50] If $t_1 = 0.066$ nm, which is close to the radius of a free carbon atom (0.0771 ~ 0.0914 nm), the $Y_1$ is derived as 5.5 TPa.[42,51,52,53,54] This uncertainty in bond thickness results in the large variety of the reported Y values.[55,56,57] Although the widely scattered values of $t_1$ and Y in measurement need to be certain, the product of Yt surprisingly approaches a constant value (0.3685 ± 0.0055 TPa·nm), as documented. The problem is that if we can develop a method allowing us to discriminate the actual Y value from the bond thickness.

For the SWCNTs, under a presumed $t_1$ value, the measured $Y_1$ varies insignificantly with the tube diameter or the tube helicity though the curvature-induced strain contributes.[58]

However, for the multi-walled (MW) CNTs, two typical trends of the change in Y values have been observed: (i) the Y value remains constant and it is almost independent of the tube diameters,[49] and (ii) the Y value increases as the number of walls ($\lambda$) is reduced.[59] Using an atomic force microscopy (AFM), Wong et al[49] measured the radius dependence of the Young's modulus of SiC nanorods and multi-wall carbon nanotubes (MWCNTs) and found that the MWCNTs are about twice as stiff as the SiC nanorods and that the strengths of the SiC nanorods are substantially greater than those found for large SiC structures (600 GPa). The Young's modulus is 610 and 660 GPa for SiC rods of 23.0 and 21.5 nm across, respectively.

In contrast, atoms at the open edge of a SWCNT could melt or coalesce prior to atoms in the tube body at temperature much lower than the melting point of the bulk graphite ($T_m$ = 3800 K). The coalescent temperature of the MWCNT increases as the number of walls increases. Coalescence of the SWCNT happens at 1073 K under energetic (1.25 MeV) electron beam irradiation and the coalescence starts at sites surrounding atomic vacancies via a zipper like mechanism.[60] The STM tip-end that is made of SWCNT starts to melt at 1593 K in ultrahigh vacuum.[61] The electron beam irradiation lowers at least the melting point by some 600 K. Annealing at 1670 - 1770 K under medium-high vacuum, or in flowing Ar and $N_2$ atmospheres, 60% SWCNTs coalesce with their neighbors.[62] Heating under an Ar flow in the temperature range of 1873 - 2273 K results in a progressive destruction of the SWCNT bundle, and then followed by the coalescence of the entire CNT bundle.[63] Coalescence starts at the edge of CNT bundles.[64] SWCNTs transform at 2473 K or higher to MWCNTs with external diameter of several nanometers. Fe-C impurity bonds can be completely removed from the CNTs at 2523 K.[65] MWCNTs are more thermally stable than the SWCNTs, and the stability of the MWCNTs increases with the number of walls.[55,66,67] On the other hand, an ordinary camera flashing[68] could burn the SWCNT at the ambient conditions, showing the higher chemical reactivity for oxidation of the SWCNT. These observations evidence consistently the lower thermal and chemical stability at raised temperatures albeit the higher mechanical strength of the CNTs at the ambient temperature.

1.2.4   Inverse Hall-Petch relationship

The mechanically strengthening with grain refinement in the size range of 100 nm or larger has traditionally been rationalized with the so-called T-unapparent Hall-Petch relationship (HPR)[69,70,71] that can be simplified in a dimensionless form being normalized by the bulk strength, P(0, T), that was measured at the same temperature and under the same conditions:

$$P(x_j, T)/P(0, T) = 1 + A x_j$$

(1)

the adjustable slope *A* is used to fit experimental data, which should be a reflection of both the intrinsic properties and the extrinsic factors such as shapes of tips. Using the dimensionless form of the normalized strength aims to minimize the contribution from artifacts due to processing conditions and crystal orientations if the measurement is conducted under the same conditions throughout the course of experiment. The $x_j = K_j^{-1/2}$, and $K_j = R_j/d$ is the dimensionless form of size, which corresponds to the number of atoms, with mean diameter or bond length d, lined along the radius of a spherical-like nanosolid or cross the thickness of a thin film. For convenience, we use both $x_j$ and $K_j$ as indicators of dimensionless form of sizes.

The pileup of dislocations at grain boundaries is traditionally envisioned as a key mechanistic process that enhances resistance to plastic flow from grain refinement. As the crystal is refined from the micrometer regime into the nanometer regime, this process invariably breaks down and the yield strength versus grain size relationship departs markedly from that seen at larger grain sizes. With further grain refinement, the yield stress peaks in many cases at a mean grain size in the order of 10 nm or so. A further decrease in grain size can cause softening of the solid, instead, and then the HPR slope turns from positive to negative at a critical size, which is often called as the inverse Hall-Petch relationship (IHPR).[72,73,74,75,76,77]

There is a concerted global effort underway using a combination of novel processing routes, experimental measurements, and large-scale computations to develop deeper insights into these phenomena. It has been suggested that the grain boundaries consisting of lower coordinated atoms contribute to the grain boundary activities.[78] The strength maximum at a grain size of 10 ~ 15 nm for Cu nanosolid is attributed to a switch in the microscopic deformation mechanism from dislocation-mediated plasticity in the coarse-grain interior to the grain boundary sliding in the nanocrystalline regime.[79] A significant portion of atoms resides in the grain boundaries and the plastic flow of the grain boundary region is responsible for the unique characteristics displayed by such materials.[80] In the HPR regime, crystallographic slips in the grain interiors govern the plastic behavior of the polycrystallite; while in the IHPR regime, grain boundaries dominate the plastic behavior. During the transition, both grain interiors and grain boundaries contribute competitively. The slope in the HPR is suggested to be proportional to the work required to eject dislocations from grain boundaries.[81] Molecular dynamics simulations suggest that the IHPR arises from sliding-accommodated grain boundary diffusion creep.[82] The critical size is suggested to depend strongly on the stacking-fault energy and the magnitude of the applied stress.[83] Unfortunately, an analytical expression for the IHPR transition was absent until the work conducted by Zhao et al.[84] who first modified the HPR by introducing the activation energy that can be related directly to the melting point suppression, $T_m(x_j) \propto E_A(x_j)$,[85] to the slope A.

1.3    Challenges

Overwhelming efforts have been exerted, primarily in regards to nanosolid growth, characterization, and function, whereas consistent insight into the origin behind the unusual behavior of the nanostructures is highly desirable. For a single phenomenon, there are often numerous theories discussing from various perspectives. Unifying all observations in a comprehensive yet straightforward way is a high challenge. For instances, the mechanisms behind the Y value enhancement and the $T_m$ suppression of the CNTs and the certainty in the wall thickness of the SWCNTs are still puzzles, although the atoms that surround defects or are located at the tip ends or at the surface are expected to play unusual yet unclear roles in dominating the mechanical and thermal behavior of the CNTs.[86]

Although there is a growing body of experimental evidence pointing to such unusual IHPR deformation responses in the nanometer regime, the underlying mechanisms are yet poorly known. As pointed out by Kumar et al.,[87] the physical origin of the IHPR transition has been a long-standing puzzle and the factors that dominate the critical size at which the HPR transits are far from clear.

Disagreement between theory and measurement still exists regarding the breaking limit of Au-Au distance in an impurity free MC and controllable formation of a MC of any metals is not very often. It is even puzzling why the harder surface skins melt easier compared with the bulk core.

One needs not only to understand the performance but also to know the origin, the trend, and the limitation of the changes and the interdependence of various properties in order to predict and control the process for fabricating materials and devices.

1.4    Objectives

The objective of this report is to review the recent progress in understanding the mechanism for the unusual mechanical and thermal behavior of systems from a flat surface, a monatomic chain, a CNT, and a nanosolid with analytical solutions that have allowed us to determine directly the length, strength, extensibility, breaking limit, specific heat and melting point of a single bond in MC. Matching the calculated Au-Au stretching limit to all the insofar-measured values at various temperatures indicates that the scattered data arises from thermal and mechanical fluctuations and that the tendency of MC formation depends uniquely on the separation of $T_m$ and T. With the known product of Young's modulus and wall (bond) thickness ($Yt \cong 0.3685$ TPa·nm) and the known temperature of tip-end melting ($T_m = 1593$ K) as well as their functional dependence on atomic coordination and bonding energy allow us to determine the dimension and strength of a C-C bond in a SWCNT by solving a group of simple equations, to advance a consistent understanding of the mechanical strength and the chemical and thermal stability of the CNTs.

An extension of the analytical solutions for an MC and a surface[88] has led to an analytical, uniscale form for quantitative information about the unusual mechanic behavior of a solid over the whole range of sizes, which provides consistent insight into the factors dominating the critical size for HPR transition and the critical temperature for solid-semisolid and semisolid-liquid transitions at a given temperature. Agreement also shows that the current model based on BOLS correlation is consistent with that derived from the criteria for melting of Lindermann[89] and Born[90] and their derivatives.

II    Principle

2.1    BOLS correlation

It is worth emphasizing that the termination of the lattice periodicity in the surface normal has two effects. One is the reduction of the coordination numbers (CNs) of the surface atoms and the other is the creation of the surface potential barrier. The potential barrier confines electrons moving inside the solid.[91] According to Goldschmidt[92] and Pauling,[93] if the CN of an atom is reduced, the radius of the lower coordinated atom would shrink spontaneously disregarding the nature of the specific chemical bond or the dimension or the structural phase of the solid.[94] It has been found[95] recently that high-density and ordered atomic layer forms at the free surfaces of liquid Sn, Hg, Ga, and In with a 10% contraction of the spacing between the first and second atomic surface layers, relative to that of subsequent layers. As impurity has induced 8% bond contraction around the impurity (acceptor dopant As) at the Te sublattice in CdTe has also been observed using EXAFS (extended X-ray absorption fine structure) and XANES (X-ray absorption near edge spectroscopy).[96] The finding of impurity-induced bond contraction could provide important impact to an atomic scale understanding of the bond in a junction interface that has been puzzled for decades. The discovery of bond contraction at liquid surface could provide a pertinent mechanism for a liquid drop formation and sustention. Therefore, the BOLS correlation can be extended to liquid and junction interfaces such as twin grain boundaries where atomic CN has little change.

The CN reduction induced spontaneous bond contraction is associated with bond strength gain or bond stiffening.[97] These sequential events have formed the subject of the intensively verified BOLS correlation mechanism that is numerically expressed as:[98]

$$\begin{cases} c_i(z_i) &= 2\{1+\exp[(12-z_i)/(8z_i)]\}^{-1} & (bond-contraction-coefficient) \\ E_i &= c_i^{-m} E_b & \begin{pmatrix} Single-bond-energy \\ Binding-energy-density \end{pmatrix} \\ E_{C,i} &= z_i E_i = z_{ib} c_i^{-m} E_b & (Atomic-cohesive-energy) \end{cases}$$

(2)

Subscript i and b denote an atom in the ith atomic layer and in the bulk. The i is counted from the outermost surface to the central of the solid up to three, as no bond order loss occurs for i > 3. The $c_i$, being the bond contraction coefficient, varies with the effective atomic CN($z_i$). The index *m*, however, is an indicator for bond nature. Exercises revealed that for elemental metals, *m* = 1, for alloys or compounds, *m* ~ 4, and for carbon and silicon, the *m* has been optimized to be 2.56[99] and 4.88,[100] respectively. The term $z_{ib} c_i^{-m} = E_{C,i}/E_{C,b} = z_i/z_b \times E_i/E_b$ is the dimensionless form of atomic cohesive energy being normalized by the bulk value. The $z_i$ also varies with the particle size due to the change of the surface curvature:

$$z_1 = \begin{cases} 4(1-0.75/K_j) & curved-surface \\ 4 & flat-surface \end{cases}$$

(3)

Generally, $z_2 = 6$ and $z_3 = 8$ or 12 would be sufficient.

Figure 1a shows the CN dependence of the reduced bond length, $c_i(z_i)$.[91] The solid curve formulates the bond order-length premise of Goldschmidt and Feibelman[101] indicating that the atomic radius contracts by 30%, 12%, 4%, and 3% if the CN of the atom is reduced from 12 to 2, 4, 6, and 8, respectively. Figure 1b shows the BOLS correlation using the pair interatomic potential, u(r). As the equilibrium atomic distance (bond length) is considered, no particular form of the u(r) is preferred. When the CN of an atom is reduced, the remaining bonds will contract from one (unit in d) to $c_i$, and the bond energy will increase in magnitude from one (unit in $E_b(T = 0)$) to $c_i^{-m}$. As a consequence of the BOLS correlation, densification happens to mass, charge, and energy at sites close to a surface or sites surrounding a structural defect, as illustrated in Figure 1c. The local densification of binding energy near the defects due to deepened trapping potential wells, may provide a possible atomistic mechanism for the "pinning effect" of atomic dislocations and atomic voids, of which energy density rise could substantially strengthen the sites surrounding the defects.[102]

Figure 1 (link)
Formulation (solid curve) of the bond order dependence of bond length derived from the Goldshmidt's premise (CN ≥ 4)[92] and Feibelman's notation.[101] (b) Consequence of bond contraction on the binding energy. CN imperfection causes the remaining bonds of the lower coordinated atom to contract from one unit (in d) to $c_i$ and the cohesive energy per coordinate increases from one unit (in $E_b$) to $c_i^{-m}$. Separation between $E_i(T)$ and $E_i(0)$ is the thermal vibration energy. Separation between $E_i(T_{m,i})$ and $E_i(T)$, or $\eta_{1i}(T_{m,i} - T)$, corresponds to energy required for melting, which contributes to the mechanical strength. $T_{m,i}$ is the melting point, which is proportional to atomic cohesive energy, $E_{C,i}$. $\eta_{1i}$ is the specific heat per bond and $\eta_{2i}$ is $1/z_i$ fold of energy required for evaporating an atom in molten state. (c) The potential energy of a nanosolid with densification of charges, mass, and energy in the surface skin.

The BOLS correlation premise has been widely applied to predicting the tunability of a variety of properties of a nanosolid. Atomic CN imperfection contributes not only to the atomic cohesive energy of the specific discrete atom but also to the energy density in the surface skin because of the potential well deepening. Atomic cohesive energy determines the thermodynamic behavior of the system such as critical temperatures for phase transition and solid-liquid transition,[9] as well as the activation energy for atomic diffusion, atomic dislocation, and chemical reaction.[103] The binding energy density perturbs the Hamiltonian of an extended solid that determines the entire band structure such as the band gap expansion,[91] core-level shifts,[97] and bandwidth modification of a nanosolid. The increased energy density in the relaxed surface region enhances the surface stress and the Young's modulus in the surface skin.[104] Therefore, all the measurable quantities of a nanosolid can be predicted given their functional dependence on bond length, bond strength, densities of mass, charge and energy, or atomic cohesive energy. Most importantly, miniaturizing a solid provides us with a new freedom that not only allows us to tune the physical properties of a solid in a predictable way but also allows us to get quantitative information, such as the energy levels of an isolated atom[97] and the frequency of vibration for an isolated Si-Si dimer.[105] Extension to the lower end of the size limit has led to the quantification of the strength, extensibility, and thermal stability of a single bond in MCs and the Young's modulus of a C-C bond in CNTs.[99]

## 2.2 Scaling relation

Generally, the mean relative change of a measurable quantity of a nanosolid containing $N_j$ atoms, with dimension $K_j$, can be expressed as $Q(K_j)$; and as $Q(\infty)$ for the same solid without contribution from atomic CN-imperfection. The correlation between the $Q(K_j)$ and $Q(\infty) = N_j q_0$ and the relative change of Q due to CN imperfection is given as:[97]

$$\begin{aligned} Q(K_j) &= N_j q_0 + N_s(q_s - q_0) \\ \frac{\Delta Q(K_j)}{Q(\infty)} &= \frac{Q(K_j) - Q(\infty)}{Q(\infty)} = \frac{N_s}{N_j}\left(\frac{q_s}{q_0} - 1\right) \\ &= \sum_{i \leq 3} \gamma_{ij}(\Delta q_i / q_0) = \Delta_{qj} \end{aligned}$$

(4)

The $q_0$ and $q_s$ correspond to the local density of $Q$ inside the bulk and in the surface region, respectively. $N_s = \sum N_i$ is the number of atoms in the surface skin composed of 2 ~ 3 atomic shells. The weighting factor, $\gamma_{ij}$, represents the geometrical contributions from dimension ($K_j$, L) and dimensionality ($\tau$) of the solid, which determines the magnitude of change. The quantity $\Delta q_i/q_0$ is the origin of change. It is easy to derive the volume or number ratio of a certain atomic layer, denoted i, to that of the entire solid as:

$$\gamma_{ij} = \frac{N_i}{N_j} = \frac{V_i}{V_j} = \frac{R_{i,out}^\tau - R_{i,in}^\tau}{R_{K_j,out}^\tau - R_{L,in}^\tau} \cong \begin{cases} \frac{\tau c_i}{K_j}, & K_j > \tau \\ 1, & K_j \leq \tau \end{cases}$$

(5)

where $\tau$ corresponds to the dimensionality of a thin plate ($\tau = 1$, and monatomic chain as well), a rod ($\tau = 2$) and a spherical dot ($\tau = 3$) of any size. *L* is the number of atomic layers without being occupied by atoms in a hollow structure. For a solid system, $L = 0$; while for a hollow sphere or a hollow tube, $L < K_j$. For a hollow system, the $\gamma_{ij}$ should count both external and internal sides of the hollow system. $d_i = R_{i,out} - R_{i,in} = c_i d$ is the thickness of the ith atomic shell. The $\sum_{i \leq 3} \gamma_{ij}$ drops in a $K_j^{-1}$ fashion from unity to infinitely small when the solid dimension grows from atomic level to infinitely large. At $K_j < 3$, performance of surface atoms will dominate because at the smallest size $\gamma_1$ approaches unity. At $K_j = 0.75$, the solid will degenerate into an isolated atom. For a spherical dot at the lower end of the size limit, $K_j = 1.5$ ($K_j d = 0.43$ nm for an Au spherical dot example), $\gamma_{1j} = 1$, $\gamma_{2j} = \gamma_{3j} = 0$, and $z_1 = 2$, which is identical in situation to an atom in a monatomic chain despite the geometrical orientation of the two interatomic bonds. Actually, the bond orientation is not involved in the modeling iteration. Therefore, the performance of an atom in the smallest nanosolid is a mimic of an atom in an MC of the same element without presence of external stimulus such as stretching or heating. At the lower end of the size limit, the property change of a nanosolid relates directly to the behavior of a single bond, which forms the starting point of the "bottom up" approach. The definition of dimensionality herein differs from convention in transport considerations in which a nano-sphere is defined as zero-dimension (quantum dot), a rod as one dimension (quantum wire), and a plate two dimension (quantum well).

Generally, experimentally observed size-and-shape dependence of a detectable quantity follows a scaling relation indicating that the size-induced property generally changes with the inverse of solid size. Equilibrating the scaling law with eq (4), one has:

$$Q(K_j) - Q(\infty) = \begin{cases} bK_j^{-1} & (measurement) \\ Q(\infty) \times \Delta_{qj} & (theory) \end{cases} \quad (6)$$

Matching theory to measurement leads to the relation, $b \equiv Q(\infty) \times \Delta_{qj} \times K_j \cong$ constant. The physical origin of the slope $b$ is the focus of various modeling pursues. The $\Delta_{qj} \propto K_j^{-1}$ (or $x_j^2$) varies simply with the $\gamma_{ij}(\tau, K_j, c_i)$. Therefore, the scaling relation covers a solid of any shape over the whole range of sizes. If the functional dependence of $q(z_i, c_i, m)$ on the atomic CN, bond length, and bond energy is given, the shape and size dependence of the Q can readily be predicted. For instances, the size induce perturbation to bond length, $d(x_j)$, and the size and bond nature dependent melting point, $T_m(x_j, m)$, crystal binding energy, $E_H(x_j)$, of a nanosolid have been given as:[106]

$$\begin{cases} \Delta d(x_j)/d(0) = \sum_{i \leq 3} \gamma_i(x_j) \times (c_i - 1) = \Delta_d(x_j); & (bond-length) \\ \Delta T_m(x_j, m)/T_m(0) = \sum_{i \leq 3} \gamma_i(x_j) \times (z_{ib} c_i^{-m} - 1) = \Delta_C(x_j, m); & (cohesive-energy) \\ \Delta E_H(x_j, m)/E_H(0) = \sum_{i \leq 3} \gamma_i(x_j) \times (c_i^{-m} - 1) = \Delta_H(x_j, m); & (Hamiltonain) \end{cases} \quad (7)$$

The sum is carried out in a way from bond-by-bond, atom-by-atom, shell-by-shell, to particle-by-particle in a system composed of weakly linked nanosolids, which adequately represents the inhomogeneous features of a shell-structured substance over the whole range of sizes. If one counts atom-by-atom, $\gamma_{ij}$ exhibits quantized features. For quantities such as the density of charge and energy, one has to consider the continuum form of $\gamma_{ij}$ by calculating the volumes of different atomic shells in the surface skins.

2.3    Analytical solutions

2.3.1  Atomic chain bonding
- Energy for melting and breaking

One needs to note in Figure 1b that the melting point of the bond between the pairing atoms, $T_{m,i}$, is proportional to the cohesive energy, $T_{m,i} \propto z_i E_i$, of the ith atom with $z_i$ coordinate,[85] and that the energy contributing to the mechanical strength at T is the separation between $E_i(T_{m,i})$ and $E_i(T)$, or $\eta_{1i}(T_{m,i} - T)$, as the molten phase is extremely soft and highly compressible.[107] The operating temperature T is critical. $\eta_{1i}$ is the specific heat per coordinate. The term of $\eta_{1i}(T_{m,i} - T)$, being a portion of the bond energy, $E_b$, is also $1/z_i$ fold thermal energy required for melting the specific atom by raising the temperature from T to $T_{m,i}$. Any means that could either raise the T or suppress the $T_{m,i}$ will lower the mechanical strength of the bond. The only way to suppress the $T_{m,i}$ is to reduce the atomic CN, as $T_{m,i}$ is proportional to the product of $z_i$ and $E_i$.

The mechanical yield strength is the strain-induced energy deviation that is proportional to internal energy density being the sum of bond energy per unit volume containing $N_i$ atoms.[108] If one wants to melt an atom with $z_i$ coordinates by raising the temperature from T to $T_{m,i}$, one needs to provide energy (see Figure 1b):

$$z_i[E_i(T_{m,i}) - E_i(T)] = z_i \eta_{1i}(T_{m,i} - T) \tag{8}$$

However, breaking a bond mechanically at a temperature T needs energy that equals to the thermal energy:

$$0 - E_i(T) = \eta_{1i}(T_{m,i} - T) + \eta_{2i}$$
$$= c_i^{-m}[\eta_{1b}(T_{m,b} - T) + \eta_{2b}] \tag{9}$$

Ideally, the slope $\eta_{1i}$ is an equivalent of specific heat per coordinate. The constant $\eta_{2i}$ represents $1/z_i$ fold energy that is required for evaporating a molten atom in the MC. $\eta_{1i}$ and $\eta_{2i}$ can be determined with the known $c_i^{-m}$ and the known values of $\eta_{1b}$ and $\eta_{2b}$ that vary with crystal structures, as listed in Table 1.

Table 1
Correlation between the bond energy and the melting point of various structures. $E_b = \eta_{1b} T_m + \eta_{2b}$ [109].

|  | fcc | bcc | Diamond structure |
|---|---|---|---|
| $\eta_{1b}$ ($10^{-4}$ eV/K) | 5.542 | 5.919 | 5.736 |
| $\eta_{2b}$ (eV) | -0.24 | 0.0364 | 1.29 |

- Strength and extensibility

Considering the contribution from heating, the strength and compressibility (under compressive stress) or extensibility (under tensile stress) at a given temperature can be expressed by:

$$P_i(z_i, T) = -\left.\frac{\partial u(r,T)}{\partial V}\right|_{d_i, T} \sim \begin{cases} \propto \dfrac{N_i E_i(0)}{d_i^\tau} & (T \ll T_{m,i}) \\ = \dfrac{N_i \eta_{1i}(T_{m,i} - T)}{d_i^\tau} & (else) \end{cases}$$

$$\beta_i(z_i,T) = -\frac{\partial V}{V\partial P}\bigg|_T = [Y_i(z_i,T)]^{-1} = \left[-V\frac{\partial u^{2^{-1}}(r,T)}{\partial V^2}\bigg|_T\right]^{-1}$$

$$= \frac{d_i^\tau}{N_i\eta_{1i}(T_{m,i}-T)} = [P_i(z_i,T)]^{-1}$$

(10)

β is in an inverse of Young's modulus or hardness in dimension.[104] $N_j$ is the total number of bonds in $d^\tau$ volume. If calibrated with the bulk value at T = 0, the temperature dependent extensibility will be:

$$\beta_i(z_i,T)/\beta_0(z_b,0) = \{N_b d_i^\tau \eta_{1b} T_{m,b}/[N_i d^\tau \eta_{1i}(T_{m,i}-T)]\}$$

$$= \frac{\eta_{1b} d_i^\tau}{\eta_{1i} d^\tau} \times \left(\frac{T_{m,b}}{T_{m,i}-T}\right)$$

(11)

The term $\eta_{2i}$ does not come into play, as the stretching can only be made before melting.[107] One needs to note that the bond number density in the relaxed region does not change upon relaxation ($N_i = N_b$). For instance, bond relaxation never changes the bond number between the neighboring atoms in a MC ($\tau = 1$) whether it is suspended or embedded in the bulk, or the number density between the circumferencial atomic layers of a solid.

- Maximal strain

By consideration of the following effects of (i) atomic CN-imperfection-induced bond contraction, (ii) thermal expansion (with linear coefficient $\alpha$), and (iii) the temperature dependence of extensibility (with coefficient $\beta$), the distance between two nearest atoms in the interior of a MC, under mechanical (P) and thermal (T) stimuli, can be expressed as:

$$d_i(z_i,T,P) = d \times c(z_i)(1+\alpha T)[1+\beta_i(z_i,T)P]$$

or the maximal strain,

$$\frac{\Delta d_{iM}(z_i,T,P)}{d_i(z_i,T,0)} = \frac{d_{iM}(z_i,T,P)}{d_i(z_i,T,0)} - 1 = \beta_i(z_i,T)\overline{P}$$

(12)

where $d_i(z_i,T,0) = d \times c(z_i)(1+\alpha T)$ is the bond length at T without mechanical stretching. At the bulk $T_{m,b}$, the liner thermal expansion ($\alpha \times T_{m,b}$) is around 2% for most metals, which is negligibly small compared with the $c_i$. Considering the energy for mechanical rupture, as given in eq (9), we have the relation:[110]

$$\int_{d_i(z_i,T,0)}^{d_{iM}(z_i,T,P)} P dx = \overline{P}[d_{Mi}(z_i,T,P) - d_i(z_i,T,0)] = \overline{P}\Delta d_{iM}$$

$$= \eta_{1i}(T_{m,i}-T) + \eta_{2i}$$

(13)

One can approximate P to the mean $\overline{P}$, if the $d_{iM}(z_i, T, P)$ represents the breaking limit, as the integral is a constant. Combining eqs (12) and (13), one has,

$$\overline{P} = \pm\left\{\frac{\eta_{1i}(T_{m,i}-T) + \eta_{2i}}{\beta(z_i,T) \times d_i(z_i,T,0)}\right\}^{1/2}$$

(14)

For tensile stress, $\overline{P} > 0$, for compressive stress, $\overline{P} < 0$. The combination of eqs (10)-(14) yields the maximal strain of a single bond in the MC:

$$\frac{\Delta d_{iM}(z_i,T,P)}{d_i(z_i,T,0)} = \beta_i(z_i,T)\overline{P} = \left\{\frac{\beta(z_i,T)\times[\eta_{1i}(T_{m,i}-T)+\eta_{2i}]}{d_i(z_i,T,0)}\right\}^{1/2}$$

$$= \left[\frac{\eta_{1b}d_i(z_i,T,0)\times\beta_0(z_b,0)}{\eta_{1i}d\times d_i(z_i,T,0)}\times\left(\frac{T_{m,b}}{T_{m,i}-T}\right)\times[\eta_{1i}(T_{m,i}-T)+\eta_{2i}]\right]^{1/2}$$

$$= \left[\frac{\beta_0\eta_{1b}T_{m,b}}{d}\left(1+\frac{\eta_{2i}}{\eta_{1i}(T_{m,i}-T)}\right)\right]^{1/2}$$

$$\cong \left(\frac{\beta_0\eta_{1b}T_{m,b}}{d}\right)^{1/2}\exp\left\{\frac{\eta_{2i}}{2\eta_{1i}[T_{m,i}-T]}\right\} = B\times\exp\left\{\frac{\eta_{2i}}{2\eta_{1i}[T_{m,i}-T]}\right\}$$

(15)

For a metallic MC with $z_i = 2$ and $m = 1$, $T_{m,i} \propto z_{ib}c_i^{-1}E_b = T_{m,b}/4.185$. The analytical expression of the maximal strain varies in-apparently with $P$ or the strain rate but apparently with the separation of $T_{m,i} - T$. The prefactor B is material nature dependent and the $\eta_{2i}/\eta_{1i}$ ratio is crystal structure dependent. The B varies with the bulk extensibility (at 0K), bulk bond length, and the $T_{m,b}$ as well as the specific heat per bond in the bulk.

2.3.2 Surface skin

For a surface skin, we have to consider the three dimensional nature, $\tau = 3$. The normalized $P_i$ and $Y_i$ at a specific site share the commonly dimensionless form:

$$\frac{P_i(z_i,T)}{P(z_b,T)} = \frac{Y_i(z_i,T)}{Y(z_b,T)} = \begin{cases}\propto c_i^{-(m+3)} & (T \ll T_{m,i}) \\ \dfrac{\eta_{1i}d^3(T_{m,i}-T)}{\eta_{1b}d_i^3(T_{m,b}-T)} & (else)\end{cases}$$

(16)

The specific heat enhancement would dominate because $(T_{m,i}-T)/(T_{m,b}-T) < 1$. Factors that enhance the $Y_i$ and $P_i$ are the shortened bond length and the associated bond strength gain.

2.3.3 Nanosolid

The size and temperature dependence of mechanical strength and compressibility or extensibility of a nansolid can be readily obtained by substituting the size and bond nature dependent $\eta_1(x_j)$, $d(x_j)$, and $T_m(x_j, m)$ for the $\eta_{1i}$, $d_i$, and $T_{m,i}$ in eq (16) that are initially derived for a single bond and a flat surface. The derived solution is expressed as:

$$\frac{P(x_j,T)}{P(0,T)} = \frac{\eta_1(x_j)}{\eta_1(0)}\left(\frac{d(0)}{d(x_j,T)}\right)^3\times\frac{T_m(x_j,m)-T}{T_m(0)-T} = \frac{P(x_j)}{P(0)}\varphi(d,m,T)$$

(17)

The additional term $\varphi(d, m, T)$ covers contributions from bond nature, the CN imperfection induced bond contraction, the $T_m$ suppression and the temperature of operation, to the mechanical strength of a solid.

By comparing the currently derived form of eq (17) with the traditional HPR (eq (1)), one can readily find that the ratio of specific heat per bond follows the traditional T-unapparent HPR, $\eta_1(x_j)/\eta_1(0) = P(x_j,T)/P(0,T) = 1 + Ax_j$. Incorporating the activation energy, $E_A \propto T_m$,[84,85,111] into the prefactor $A$, leads to an analytical expression for the size and temperature dependent HPR:

$$\frac{P(x_j,T)}{P(0,T)} = \frac{\eta_1(x_j)}{\eta_1(0)}\left(\frac{d(0)}{d(x_j)}\right)^3 \times \frac{T_m(x_j,m)-T}{T_{m,b}-T}$$

$$= [1+A(x_j,T_m(x_j,m),m,T)] \times [1+\Delta_d(x_j)]^{-3} \times \frac{T_m(x_j,m)-T}{T_{m,b}-T}$$

(18a)

$$\text{where, } A(x_j,T_m(x_j),m,T) = f \times x_j \times \exp\left[\frac{T_m(x_j)}{T}\right]$$

(18b)

The prefactor *f* is an adjustable parameter, which should cover contribution from extrinsic factors such as impurities or defects, shapes of tips, and loading scales to the mechanical strength of a solid.

The analytical form supports for the known grain-interior – grain-boundary transition mechanism for the IHPR[78-81] and further clarifies that the competition between bond order loss ($T_m$ suppression) and bond strength (specific heat) gain dictates the mechanical behaviour at grain boundaries, which is sensitive to the temperature of operation. As the solid size is decreased, transition from the dominance of bond strength gain to the dominance of bond order loss happens at the critical size at which both mechanisms contribute competitively.

According to this solution, grain boundary is harder at temperatures far below $T_m(x_j, m)$ because of the dominance of bond strength gain, whereas at temperatures close to $T_m(x_j, m)$, the grain boundary is softer than the grain interior because of the bond order loss that lowers the barrier for atomic dislocation. If operating at a given temperature, solid size reduction lowers the $T_m(x_j, m)$. When $x_j$ approaches zero, the analytical form converges to the traditional HPR, of which the slope is clarified herein to be dominated by the term of f × exp($T_m(0)/T$) that relates to the specific heat or activation energy for atomic dislocation. For twined grains, one has to consider the interfacial strengthening that is expected to elevate both the mechanical strength and thermal stability, being the cases of nanocrystalline/amorphous composites such as nc-TiN/a-$Si_3N_4$, nc-TiN/a-$Si_3N_4$/ and nc-$TiSi_2$, nc-($Ti_{1-x}Al_x$)N/a-$Si_3N_4$, nc-TiN/$TiB_2$, nc-TiN/BN, showing hardness approaches that of diamond because of the interfacial mixing effect[112,113] and the case of high density twins in pure Cu.[114]

- Critical sizes for IHPR transition

Letting $d(Ln(P/P_0))/dx = 0$, eq (18a), we can readily find the critical size $x_C$(f, $T_m(x_j)$ T, m), at which the slope of HPR transits from positive to negative, and, hence, the factors dominating the $x_C$. For simplicity, we define $\theta(x_j) = T/T_m(x_j)$. P and $P_0$ represent $P(x_j, T)$ and $P(0, T)$, respectively. The numerical process leads to the following relation:

$$\frac{d(Ln(P/P_0))}{dx} = A(x_j,\theta(x_j),m) \times \frac{\theta(0)+2\Delta_C(x_j,m)}{\theta(0)(1+A(x_j,\theta(x_j),m))}$$

$$-\frac{6\Delta_d(x_j)}{1+\Delta_d(x_j)} + \frac{2\Delta_C(x,m)}{1+\Delta_C(x,\theta(x_j),m)-\theta(0)} = 0$$

(19)

Solution indicates that the critical size depends on the bond nature indicator *m*, $T/T_m(0)$, and the prefector *f*.

- $T_C$ for solid-semisolid and semisolid-liquid transition

We may define the critical temperatures, $T_C$, for the solid-semisolid transition and $T_m$ for semisolid-liquid transition by the relation:

$$\frac{P(x_j, T_C)}{P(0, T_C)} = \frac{1 + A(x_j, \theta(x_C), m)}{[1 + \Delta_d(x_j)]^3} \times \frac{T_m(x_j, m) - T_C}{T_{m,b} - T_C} \begin{cases} \leq 1 & (Semisolid) \\ = 0 & (Liquid) \end{cases}$$

(20)

At temperature higher than $T_C(x_j, f, m)$, the solid is softer and easily compressible compared with the bulk counterpart at the same temperature. At the melting point, $T_m(x_j, m)$, the semisolid-to-liquid transition happens associated with zero hardness and infinity compressibility. This definition is quite the same to Born's criterion indicating the absence of shear modulus at melting.[90]

III    Results and discussion

3.1    Surface relaxation and nanosolid strain

For a freestanding nanosolid, the lattice constants are often measured to contract while for a nanosolid embedded in a matrix of different materials or passivated chemically, the lattice constants may expand. For example, oxygen chemisorption could expand the first metallic interlayer by up to 10%-25% because of oxygen insertion into the neighboring atomic layers. However, the oxygen-metal bonds still contract.[115,122,116] The mean lattice constants of *Sn* and *Bi* nano-particles contract with particle size.[117] The absolute contraction of the c-axis lattice is more significant than that of the a-axis lattice but the relative strain is identical.[118] Kara and Rahman[119] found theoretically that the bond length between neighboring atoms of *Ag*, *Cu*, *Ni*, and *Fe* decrease with decreasing coordination. The lengths of the dimer bonds are shorter than the nearest-neighboring distance in their respective bulk values by 12.5% for *Ag*, 13.2% for *Cu*, 13.6% for *Ni*, and 18.6% for *Fe*, associated with bond energy enhancement by 2 ~ 3 times. The pattern of dimer bond relaxation coincides with the surface relaxations of the top layer atoms for these elemental solids with various orientations. The findings are in line with the BOLS correlation albeit the extent of relaxation and energy enhancement. As listed in Table 2, most of the bonds contract disregarding the bond nature or the phase structures, which has enormous effect on the properties of a surface. Bond expansion might happen but the system energy must be minimized, unless the relaxation is a process proceeding under external stimulus such as heating or stretching.

Table 2 Bond length relaxation for typical covalent, metallic and ionic solids and liquids and its effect on the physical properties of the corresponding solid or surface. Where $d$ and $d_1$ is the bond lengths for atoms inside the bulk and for atoms at the surface, respectively. The $c_1$ is the bond contraction coefficient.

| Bond nature | Medium | $c_1 = d_1/d$ | Effect |
|---|---|---|---|
| Covalent | Diamond {111} [120] | 0.7 | Surface energy decrease |
| Metallic | Ru[121], Co[122] and Re[123])($10\bar{1}0$) surfaces | 0.9 | |
| | Fe-W, Fe-Fe [124] | 0.9 | Atomic magnetic momentum is increased by (25~27)% [124-126] |
| | Fe(310)[125],Ni(210)[126] Al(001) [127] | 0.88 | |
| | | 0.88 | |
| | Ni, Cu, Ag, Au, Pt, and Pd dimer bond [23] | 0.85-0.9 | Cohesive energy rises by 0.3 eV per bond [127]. |
| | Ti and Zr dimer bond[101] | 0.7 | |
| | V dimer bond [101] | 0.6 | Single bond energy |
| | Ta(001)[128] | 0.9 | increases by 2 ~ 3 times.[23] |
| | Nb(001)[129] | 0.88 | 0.50 ~ 0.75 eV positive shift of Nb-3d and Ta-4f core level |
| Ionic | O-Cu(001) [130,131] | 0.88-0.96 | |
| | O-Cu(110) [131] | 0.9 | N-TiCr surface is 100% harder that the bulk. |
| | N-Ti/Cr [104] | 0.86-0.88 | |
| Liquid surface | Sn, Hg, Ga, and In [95] | 0.9 | |
| Abnormal cases | (Be, Mg)(0001) Zn, Cd and Hg dimer bond [101] | > 1.0 | No indication of effects on physical properties is yet given. |

3.2 Surface mechanics

An examination of the hardness and the Young's modulus using nanoindentation revealed that the surface of TiCrN thin films[104] reaches a maximum of 100% higher than its bulk value, as shown in Figure 2a. The same trend holds for amorphous carbon[132] and AlGaN[133] films with peaks positioned at several nanometers' depth that corresponds to the surface roughness. Solving eq (16) with the measured value of $\Delta P/P = (50-25)/25 = 1$ gives rise to $c_i$ value of 0.883 associated with m = 4. Figure 2b shows consistency between BOLS predictions and the theoretically calculated thickness dependence of Young's modulus for Ni, Cu, and Ag thin films at $T \ll T_m$.[134,135] BOLS predictions were obtained by summing Eq(16) over the top three atomic layers with $z_1 = 4$, $z_2 = 6$, and m = 1 for the pure metals. It is interesting to note that at the thinnest limit (two atomic layers), the Young's modulus of Cu is 100% higher than the bulk value, which agrees with that detected from the surface of TiCrN,[104] $\alpha_2$-TiAl and $\gamma$-TiAl films.[4] Correlation between the maximal Y values and the lattice contraction of Liu et al.[6] leads to the $c_i$ value of 94.8 % for steel, which agrees with the BOLS low-T prediction of $Y_i/Y_b = c_i^{-(m+3)}$ with m = 1 for metals.

> Figure 2 (link)
> (a) Nanoindentation hardness-depth profile for TiCrN thin film.[104] The peak shift corresponds to the surface roughness (Ra = 10 nm as confirmed by AFM). (b) Agreement between predictions and measured size dependence of the Young's modulus of *Ni*-I and *Cu*-I,[134] Ni-II, Cu-II and Ag[135] thin films.

However, the high $P_i/P > 2$ values for Ag, Ni, Cu, Al thin films[4] and amorphous carbon films[3] are beyond the BOLS expectation. The unexpected high surface hardness may be a contribution from tip shapes or loading scales as discussed already.[2,136] Structural defects also contribute to the mechanical strength of a solid because of the pinning effect.[102] For instance, the internal stress of amorphous carbon films can be modulated by changing the sizes of nanopores that are produced by the bombardment of noble gases (Ar, Kr, and Xe) during formation.[137,138] When 1 ~ 11 GPa internal stress is generated by controlling the size of the pores within which noble gases are trapped, the core level binding energy of the entrapped gases is lowered by ~ 1 eV associated with 0.05 nm expansion of the atomic distance of the entrapped gases. For Ar (Xe), the interatomic separation varies from 0.24 (0.29) nm to 0.29 (0.32) nm in the 1–11 GPa pressure range. The binding energy weakening and atomic distance expansion of the entrapped gases indicate clearly that the gas-entrapped pores expand in size and the interfacial C-C bonds contract, which contribute to the extraordinary mechanical strength of the thin films.

3.3 Monatomic chains
3.3.1 Equilibrium distance

For a spherical dot at the lower end of the size limit, $K_j = 1.5$ ($R_j = K_j d = 0.43$ nm for an Au spherical dot), $\gamma_{1j} = 1$, $\gamma_{2j} = \gamma_{3j} = 0$, and $z_1 = 2$, which is identical in situation to an atom in a MC. At $R_j = 0.43$ nm, the Au-Au distance contracts by ~30% from 0.2878 to ~ 0.2007 nm, according to the BOLS premise, which is close to the value, $0.23 \pm 0.04$ nm, measured at 4.2 K under stretching[22] and close to the calculated shorter Au-Au distance of 0.232 nm as well.[23]

3.3.2 Binding energy and thermal stability

Figure 3a shows the match between the predicted and the measured size dependence of 4f core-level shift $\Delta E_{4f}(\infty) = E_{4f}(\infty) – E_{4f}(1)$ for Au nanofilms deposited on octanedithiol,[139] $TiO_2$[140] and Pt[141] substrate, and thiol-capped Au particles.[142] Results indicate that the bond energy of an Au-Au MC is increased by $c_i^{-1} – 1 = 1/0.7 – 1 \sim 43\%$. Results imply that the Au-4f core-level drops abruptly from that of an isolated atom by a maximum of 43% upon the MC being formed and then the shift recovers in a $K_j^{-1}$ fashion to the bulk value when the solid grows from atomic scale to macroscopic size.

Calibrated with $Q(\infty) = T_m(\infty) = 1337.33$ K and by use of the same m (= 1) value used in calculating the $E_{4f}(K_j)$, we obtained the theoretical $T_m$-suppression curves for different shapes, which are compared in Figure 3b with the measured size-dependent $T_m$ of Au on W[143] and C[144] substrates, and Au particles encapsulated in Silica matrix.[145] The theoretically predicted melting curves merge at the lower end of the size limit, $K_j = 1.5$ with a ~75% suppression. Therefore, thermal rupture of the Au-Au chain occurs at ~320 K that is much lower than the bulk melting point of 1337.33 K. At low temperature, the shortened bond is twice stronger (strength = $E_i/d_i \propto c_i^{-2} = 0.7^{-2} \approx 2$), agreeing with predictions of ref 28. This means that more force is required to stretch or compress a single bond in the MC by the same length compared to the force needed to stretch the same single bond in the bulk by the same amount at very low temperature. **Table 3** tabulates the information derived from the matching to size dependence of the Au-4f level shift and melting point suppression of Au nanosolids.

Figure 3 (link)
Comparison of theory with the measured size dependence of (a) $[E_{4f}(K_j) - E_{4f}(\infty)]/[E_{4f}(\infty) - E_{4f}(1)]$ of Au (nano-dot) on different substrates with derived information as given in Table 2, and, (b) $[T_m(K_j) - T_m(\infty)]/T_m(\infty)$ of Au nanosolids on W [143] and C [144] substrates and embedded in $SiO_2$ matrix [145], showing strongly interfacial effects and dimensionality transition on the Au nanosolid melting. Melting at the lower end of size limit (K = 1.5, z = 2) corresponds to the situation of an Gold MC.

Table 3 The length and energy of the Au-Au bond in the monatomic chain and the core-level energy of an isolated Au atom obtained from decoding the size dependent $E_{4f}(K_j)$ and $T_m(K_j)$ of nanosolid Au.[97]

|  | Au/Octan | Au/TiO$_2$ | Au/Pt | Au/Thiol |
|---|---|---|---|---|
| τ (dimensionality) | 3 | 1 |  | 3 |

| $E_{4f}$(eV) | -81.504 | -81.506 | -81.504 | -81.505 |
|---|---|---|---|---|
| $\Delta E_{4f}(\infty)$(eV) | -2.866 | -2.864 | -2.866 | -2.865 |
| $d_{MC}/d$ | 0.2001/0.2878 | | | |
| $E_{MC}/E_b$ | 1.43 | | | |
| $T_{m,MC}/T_m(\infty)$ | 1/4.185 | | | |

### 3.3.3 Strain limit

In order to examine the validity of the prediction, we employed the known values of thermal expansion coefficient $\alpha = 14.7\times10^{-6}$, $T_{m,b}$ = 1337.33 K, d = 0.2878 nm to predict the maximal strain of the gold MC using eq (15). The measured breaking limits of $d_{iM}$(4 K) = 0.23 nm,[22] and the mean $d_{iM}$(300 K) = 0.35,[17,19,20] were used, which leads to the quantification of the two unknown parameters of $\beta_0$ = 0.005 GPa$^{-1}$ and $\eta_{2i}/\eta_{1i}$ = 64 K.

Noting the relation $E_i = c_i^{-1}E_b$, and hence, $\eta_{1i}T_{m,i} + \eta_{2i} = c_i^{-1}(\eta_{1b}T_{m,b} + \eta_{2b})$, with the given $\eta_{1b}$ = 0.0005542 eV/K and $\eta_{2b}$ = -0.24 eV for the fcc structures,[109] $\eta_{1i}$ = 0.00187 eV/K ~ 3.3742$\eta_{1b}$, and $\eta_{2i}$ = 0.1197 eV were obtained. The $\eta_{2b}$ < 0 in Ref [109] means that the actual energy for evaporating the molten atom is included in the term of $\eta_{1b}T_m$, and therefore, the $\eta_{1b}$ there may not represent properly the specific heat per coordinate. Accuracy of solutions gained herein is subject strictly to the $\eta_{1b}$ and $\eta_{2b}$ values and the precision of the measured $d_{iM}(T \neq 0)$ values used for calibration, as no freely adjustable parameters are involved.

Figure 4 compares the calculated curves for strain versus $T/T_m$ with the measured maximal strains for a gold MC at various temperatures. Excitingly, the theoretical curve covers all the divergent values measured at 4 K (0.23 ± 0.04 nm) and at the ambient temperatures (298 ± 6 K, 0.29 ~ 0.48 nm). The divergent data are centered at some 22 K below the MC melting point, 320 K, with 6 K fluctuation. The fluctuation may arise from environment temperature or the electron beam energy in the TEM. Finding shows clearly now that the divergent values of breaking limit measured at the ambient temperature originate from thermal and mechanical fluctuations near the $T_{m,i}$ of the MC that is 1/4.185 fold of the bulk value, and therefore, possibility of impurity meditation is excluded. As a matter of fact, there hardly exist of light elements in a TEM vacuum chamber.

Figure 4 also suggests that the dominant factor $T_{m,b}$ has slight influences on the breaking mode of a MC. The bond of a low $T_{m,b}$ specimen breaks more readily at low temperature than the bond of a high $T_{m,b}$ element; the bond of the low $T_{m,b}$ specimen is more easily extended as T approaches the $T_{m,i}$.

### 3.3.4 Criteria for MC formation

Equation (15) indicates that a metallic MC melts at a temperature that is 1/4.185 fold the bulk $T_{m,b}$. A metallic MC could be made only at temperature that is 20 ± 6K lower than its melting point, $T_{m,i}$. Therefore, if one wants to make an MC of a certain specimen at the ambient temperature (300 K), one has to work with the material whose $T_{m,b}$ satisfies (300 + 20) ×4.185 = 1343 K or around. However, an MC can hardly form at room temperature or above if the bulk $T_{m,b}$ of the specific metal is below 300 × 4.185 = 1260 K, such as Sn (505.1 K), Pb (600.6 K), and Zn (692.7 K). A Ti-MC (with $T_{m,b}$ = 1941 K) may form at ~440 K, slightly lower than its $T_{m,i} = T_{m,b}/4.185$ = 462 K. However, it is quite possible to make a specific MC by operating at a carefully calculated and controlled temperature. It is not surprising that Au is favorable for MC formation at the ambient temperature whereas Ag ($T_{m,b}$ = 1235 K), Al ($T_{m,b}$ = 933.5 K) and Cu ($T_{m,b}$ = 1356 K) are hardly possible though they could form NWs with high extensibility. Although the electron structure may need to be considered in making an MC,[146] the operating temperature would be critical in making a MC.

The high extensibility is dominated in the temperature range that should correspond to the semisolid state. This finding explains why some metallic MCs can form at the ambient temperature and some can not.

Figure 4 (link)
Temperature ($x = T/T_{m,b}$) dependence of an MC breaking limit in comparison with values for an Gold MC measured at 4 K (0.23 ± 0.04 nm) and at ambient (298 ± 6 K, 0.29 ~ 0.48 nm) indicates that the scattered data arise from thermal and mechanical fluctuations near the melting point of the Gold MC which is around 320 K. Varying the $T_{m,b}$ changes slightly the easiness of MC bond breaking at different temperatures [88].

3.4 Metallic nanowires
3.4.1 Extensibility

If the $T_{m,i}$ and the $\eta_{2i}/\eta_{1i}$ for an MC in eqs (11) and (15) are replaced with the size dependent $T_m(K_j)$ and $\eta_2(K_j)/\eta_2(K_j)$, the extensibility or compressibility and the strain limit of a defect-free nanosolid becomes:[147]

$$\frac{\Delta\beta_i(z_i, K_j, T)}{\beta_0(z_b, 0)} = \frac{\eta_{1b}c(K_j)T_{m,b}}{\eta_1(K_j)[T_m(K_j) - T]} - 1$$

$$\frac{\Delta d_M(K_j, T, P)}{d(K_j, T, 0)} \cong B \exp\left\{\frac{\eta_2(K_j)}{2\eta_1(K_j)[T_m(K_j) - T]}\right\}$$

(21)

Counter plots in Figure 5 illustrate the size and $T/T_m$ dependent extensibility and the maximal strain of impurity-free Au-NWs with the parameters determined for the gold MC. The ratio $\eta_{1b}/\eta_1(K_j) = 1+4/(1+\exp((K_j-1.5)/20))$ is assumed to change from 0.0005542/0.00187 ~ 1/3.374, (at $K_j = 1.5$ as just determined) to 1 (at $K_j = \infty$) gradually. For an NW, $\tau = 2$. Noting that the $c(K_j)$ drops from 1 to 0.7, and the $T_m(K_j)/T_m(\infty)$ drops from 1 to 1/4.185 when a NW of infinite size shrinks to an MC. Eq. (21) indicates that the extensibility enhancement happens ($\Delta\beta_i > 0$) when $[T_m(K_j) - T] < \eta_{1b}c(K_j)T_{m,b}/\eta_1(K_j)$, otherwise, the extensibility is lower than the bulk value. When the $T_m(K_j)$ approaches T, the extensibility increases exponentially until $T \sim T_m(K_j)$.

Measurements have shown that the detectable maximal strain of a suspended impurity-free gold MC bond is less than 100% ((0.48 - 0.288)/0.288), which is much lower than the detected strain ($10^3$) of a nanograined Cu and aluminum NWs that could form at room temperature or the subambient.[148,149,150] Therefore, bond stretching discussed herein is not the factor dominating the high extensibility of a NW. The mechanism of atomic gliding dislocations and grain boundary movement should be the dominant ones.[16148,149,150] However, understanding further clarifies that the barrier or activation energy for atomic dislocation and diffusion of the lower coordinated atoms at the grain boundaries is lower than that of a fully coordinated atom in the bulk, as these activities are subject to atomic cohesion that drops with atomic CN.

Figure 5 (linka, Link b)
Illustrative counter plots for the $K_j$ and $T/T_{m,b}$ dependent (a) extensibility and (b) maximal strain of defect-free Au-NWs. The extensibility and the maximal strain increase rapidly when T approaches to $T_m(K_j)$ that drops with $K_j$.

### 3.4.2 Breaking modes

Figure 6 showing the atomic CN and temperature dependence of maximal bond strain indicates that the melting of a nanosolid starts from the first surface shell (z = 4) and then the next (z = 6) when the temperature is raised. This agrees with the 'liquid skin nucleation and growth' fashion[151,152,153,154,155] and the 'surface phonon instability' models[156,157] for nanosolid melting. At temperatures close to the $T_{m,1}$ of the surface, the maximal strain and the extensibility of the surface layer approach to infinity. At these temperatures or above, bond breaking under tension should start from the NW interior, because the inner bonds firstly reach their strain limits that are lower than that of the surface skin at the same temperature.

However, at temperatures far below the $T_{m,1}$, bond breaking may start from the outermost atomic shell and the nanosolid manifests brittle characteristics, as the shortened surface bonds break first. At very low temperatures, the theoretically allowed maximal strains for the surface shells should be constant. If one deforms the entire nanowire by *xd* per unit *d* length, the strain of the bonds in the respective shells will be $\varepsilon_i = x/c_i$. Because $c_1 < c_2 < c_3$, then the actual strain is $\varepsilon_3 < \varepsilon_2 < \varepsilon_1$, which indicates that the surface shell bond breaks first at low T. Therefore, the breaking mode of a nanowire at low T is expected to be opposite to that at higher T. At low T surface bond breaks first while at high T surface bond breaks last.

It has been measured using a "nanostressing stage" located within a scanning electron microscope at room temperature that the MWCNTs break in the outermost shell.[158] This failure mode called as 'sword-in-sheath' agrees with the expectation of the current approach as the operating temperature is far below the tube melting point that is 1600 K or higher.[99] The helical multi-shell gold nanowires[159] become thinner and thinner without breaking the outer-shell atomic bond under tension at room temperature,[160] as the gold NW breaking starts from the inner shell according to the current understanding.

Figure 6 (link)
Temperature and CN (z) -dependence of the Au-Au bond maximal strain shows the order of melting at a curved surface and implying the breaking mode of a nanowire at different temperature range, see text.

### 3.5 Nanobeams
### 3.5.1 C-C bond in CNT

The striking difference between the bulk carbon and a SWCNT is that the effective atomic CN (or $z_i$, i = 1 for a SWCNT) of a C atom reduces from a CN of 12 (diamond) to a CN of 3 upon SWCNT formation. For an atom that is surrounding a defect or is located at the tip end of the open edge, the CN is 2. The effective atomic CN of a C atom in the diamond bulk is always 12 rather than four as the diamond is an interlock of two fcc primary unit cells. Comparing the covalent bond length of diamond (0.154 nm) with that of graphite (0.142 nm), the effective atomic CN of carbon in graphite is derived to be 5.5 according to the BOLS iteration (eq (2)).

Considering the fact of $T << T_{m,b}$, the functional dependence of the $Y_i$ and $T_{m,i}$ on the atomic CN and bond energy at equilibrium atomic separation is:

$$\begin{cases} Y_i = v \dfrac{\partial P}{\partial v}\bigg|_{r=d_i} = v \dfrac{\partial^2 u(r)}{\partial v^2}\bigg|_{r=d_i} \propto n_i E_i = d_i^{-\tau} E_i = c_i^{-(m+2)} Y_b \\ T_{m,i} \propto z_i E_i \end{cases}$$

(22)

For a SWCNT that is formed by rolling a graphite sheet, $\tau = 2$, as $n_1 \propto d_1^{-2}$ is the bond number per unit area, which is independent of the wall thickness. Obviously, no other argument could change the $Y_i$ value unless the bond length was shortened and/or the single bond energy was increased. At the melting point, the bonds of the lower-coordinated atom will be thermally loosened and the atom will coalesce with its neighbors.

The known values of $(Yt)_1 = 0.3685$ TPa·nm for $z = 3$, and the known $T_{m,1}$ (for $z = 2$ at the tip end) value (= 1593 K) for a SWCNT and the known bulk values for carbon ($T_{m,b}$ = 3800 K, $Y_b = 1.02$ TPa for $z = 12$) should satisfy the relation:

$$\begin{cases} \dfrac{T_{m,2}}{T_{m,12}} = z_{2,12} c_i(2)^{-m} = c_i(2)^{-m}/6 & (tip-open-edge) \\ \dfrac{T_{m,i}(3)}{T_{m,i}(2)} = z_{3,2}\left(\dfrac{c_i(3)}{c_i(2)}\right)^{-m} = \dfrac{3}{2}\left(\dfrac{c_i(3)}{c_i(2)}\right)^{-m} & (wall-tip-relation) \\ \dfrac{(Yt)_i}{Y_b t_i} = c_i(3)^{-(2+m)} & (CNT-wall) \end{cases}$$

(23)

Subscript numbers indicate the CN values. The value of $c_i(z)$ is given in eq (2). These simple relations give immediate solutions of $m = 2.5585$, $t_i = 0.142$ nm and the tube-wall melting point, $T_{m,i}(3) = 1605$ K. Furthermore, the activation energy for chemical reaction is also a portion of the atomic cohesive energy. Therefore, the chemical stability of the lower-coordinated atoms is lower than the bulk values, which may explain why the CNT could burn using an ordinary camera flash under the ambient conditions.

The accuracy of the numerical solutions is subject to the measured input of $T_{m,1}(2)$ and the product of Yt. Errors in measurement or structural defects of the CNT may affect the accuracy of the solutions; however, they never determine the nature of the observations. Variation of any of the input (bold figures in Table 4) leads to solutions that are physically forbidden. For example, replacing the quoted Yt value with Y = 0.8 TPa (disregarding the thickness $t_1$) gives $m = -3.2$. The corresponding $E_1/E_b = c_i(3)^{-m} = 0.52$, and $T_{m,1}(3) = z_{1b} \times c_i(3)^{-m} \times T_{m,b} = 493$ K are unacceptable because the $T_{m,1}(3)$ value is much lower than any reported values. Assuming the SWCNT tip-end melting point to be $T_m(2) \geq 1620$ K gives a m value larger that 2.605 and $T_m(3) \leq 1620$ K, which is not the case of observation: tip-end melts easier. If a value of $m = 4$ is taken, $T_{m,1}(2) = 2674$ K and $T_{m,1}(3) = 2153$ K, which are much higher than the measured values despite the order of melting. Therefore, the solution with the measured input Yt and $T_{m,1}(2)$ values is unique and, hence, the quoted Yt and $T_{m,1}(2)$ values are essentially true for the SWCNT.

Table 4 Comparison of the calculation results with various input parameters (bold figures) proving that the obtained solution is unique and the quoted data represent the true values.[99]

| Yt (TPa·nm) or Y(TPa) | $(Yt)_1 = 0.3685$ Y = 2.595 | ***Y = 0.8*** | $(Yt)_1 = 0.3685$ | |
|---|---|---|---|---|
| Tip-end $T_{m,1}(z = 2)$ (K) | 1593 | - | 2153 | **≥1620** |
| Tube wall $T_{m,1}(z = 3)$ (K) | 1605 | 493 | 2674 | ≤1620 |
| m | **2.5585** | 2.5585 | **4** | 2.6050 |
| Bond thickness $t_1$(nm) | 0.142 | - | - | - |
| Bond length $d_1$(nm) | 0.116 | - | - | - |

| Bond energy $E_1/E_b$ | 1.68 | 0.53 | - | - |
|---|---|---|---|---|
| Remarks | Acceptable | Forbidden | | |

### 3.5.2 Nanobeam stiffness

For a hollow tube, the surface to volume ratio is:

$$\gamma_{ij} = \frac{\left(D_{out,i}^2 - D_{in,i}^2\right)_M + \left(D_{out,i}^2 - D_{in,i}^2\right)_m}{D_{Mj}^2 - D_m^2} \propto \frac{1}{\lambda_j} \tag{24}$$

Compared with a nanosolid, the $\gamma_{ij}$ for a MWCNT considers both the inner and outer skins. $D_M$ ($D_M = 2K_j d$) is the outer diameter and $D_m$ ($D_m = 2(K_j-\lambda_j)d$) is the inner hollow diameter. $K_j d$ is the radius of the tube and $\lambda_j d$ is the wall thickness for a MWCNT. $D_{out,i}$ and $D_{in,i}$ correspond to the outer and inner radius of the *i*th atomic layer ($D_{out,i} - D_{in,i} = d_i$). The ratio $\gamma_S = \sum_{i \leq 3} \gamma_{ij}$ decreases in an inverse fashion ($\lambda_j^{-1}$) from unity to infinitely small, when $\lambda_j$ grows from unity to infinity. Therefore, it is not surprising that, for a solid rod or a MWCNT with $K_j \leq \lambda_j$, the overall quantity change, $\Delta Q(\lambda_j)/Q$, varies with the inverse radius ($1/K_j d$) and the $\Delta Q(\lambda_j)/Q$ value differs from the corresponding bulk value ($\Delta Q(\lambda_j)/Q = 0$). For a hollow MWCNT with constant $\lambda_j$, the $\Delta Q(\lambda_j)/Q$ should vary with the diameter of the MWCNT insignificantly. These predictions agree well with the observed mechanical strength of nanobeams.[49] The broad range of measured values should be more due to the fluctuation in the number of walls of the MWCNTs than to the error in measurement.

### 3.6 Nanosolid
### 3.6.1 Strain induced stiffness

Combining the pair distribution function (PDF) derived from wide-angle x-ray scattering and extended x-ray absorption fine structure (EXAFS) analyses, Gilbert et al.[161] found the structural coherence loss of 3.4 ± 0.3 -nm-diameter ZnS nanoparticles happens over distances of 2 nm rather than the full size of the examined nanograin, 3.4 nm. The PDF for real ZnS nanoparticles is distinct from that of ideal ZnS nanoparticles in the following respects, as shown in Figure 7:[161] (i) The first-shell PDF peak intensity is lower compared with that for the ideal case. (ii) PDF peak intensities at higher correlation distances diminish more rapidly than the ideal nanoparticle. (iii) PDF peak widths are broader in the real nanoparticle. (iv) PDF peak positions are shifted closer to the reference atom. The shift is more apparent at r = 1.0 nm and 1.4 nm (shortened by 0.008 and 0.02 nm, respectively), indicating a contraction of mean bond length of the nanoparticle. The vibration frequency of ZnS nanoparticles is estimated to increase from the bulk value of 7.12 ± 1.2 to 11.6 ± 0.4 THz, implying bond stiffening.

Figure 7 (Link) Comparison of the pair distribution function of ZnS bulk solid, calculation for ideal nanosolid and the measurement for real nanosolid show the cohesive length loss of nanosolid. [161]

The findings of straining induced stiffening for ZnS nanosolid agree appreciably well with the BOLS prediction suggesting that the PDF intensity weakening or diminishing results from number loss of high-order CN atoms. The PDF peak shifting and broadening arise from the inhomogeneous bond contraction in the outermost two or three atomic layers and the non-uniformity of nanosolid sizes. As the XRD and the EXAFS collect statistic information from numerous nanosolids, one could hardly recognize the structure distortion arises either from the surface region or from the interior of the nanosolids. However, the diameter difference of (3.4 - 2.0) 1.4 nm corresponds to the scale of thickness of the outermost atomic capping and surface layers[162] (3 × 2× 0.255 nm) of which atoms are subject to CN imperfection. Compared with the PDF of an amorphous solid of which the structure coherence extends only to a couple of atomic distances, the detected PDF coincides with the core size of the measured crystalline nanosolid. Therefore, the surface layers dominate the bond length distortion. The strengthening of the shortened bonds is responsible for the stiffening of the entire nanosolid. The difference between a solid composed of nanoparticles and a solid in amorphous state is the distribution of atoms with CN imperfection. For the former, lower-coordinated atoms are located at the surface; for the latter, they distribute randomly inside the solid and the distribution is sensitive to processing conditions. In collecting statistic information, the low-CN atoms contribute identically irrespective of their locations. It is anticipated that the PDF correlation length, or core size, increases with solid dimension and further verification would be helpful.

3.6.2 Inverse Hall-Petch relationship

By incorporation of the value of $\eta_1(x_j)/\eta_{1b} = 0.00187/0.0005542 = 3.3742$ for an impurity-free gold MC into eq (18a), we may estimate the value of $f$ for the gold MC with the parameters of $z_i = 2$, $K_j = 1.5$, $x_j = 1/\sqrt{K_j} = 1.223$, $m = 1$, $T = 298$ K, and $T_m(x_j) = 1337.33/4.1837$:

$$f = \left[\frac{\eta_1(x_j)}{\eta_1(0)} - 1\right] \bigg/ \left[x_j \times \exp\left[\frac{T_m(x_j)}{T}\right]\right] = 2.3742 \bigg/ \left\{1.223 \times \exp\left[\frac{1337.33}{298 \times 4.1837}\right]\right\} = 0.663$$

(25)

Experimental conditions such as the external stress or strain rate or structural defects will contribute to the mechanical strength through the prefactor $f$ that is intrinsic for the impurity-free gold. The $f$ value may change upon bond nature alteration, being the cases of Si and $TiO_2$ that will be shown shortly.

Further calculations using eq (18a) were performed on some nanosolids with known d and $T_m$ values as input and $x_C$ as output shown in Table 5. All the nanograins were taken as in spherical shape. The prefector $f$ is adjusted under the constraint that the slope should match the observations and intercept at the positive side of the vertical axis. A negative intercept in measurement is physically unacceptable. The theoretical curves were normalized with the calculated peak values at the transition point, $x_C$, and the experimental data measured at room temperature were normalized with the measured maximum $P_M$.

Table 5
Prediction of the critical sizes for IHPR transition with f = 0.5, 0.663 and m = 1 otherwise as indicated. The predicted critical sizes agree with measurement. Changing the f value only affects the $D_C$ (= $K_C d$ = $x_C^{-2} d$) of a material with $T_m$ below 1000 K.[163]

| Element | d/nm | $T_m$/K | $P_M$ | Measured $D_C$/nm | Predicted $D_C$/nm (f = 0.5; 0.663) |
|---|---|---|---|---|---|
| Mg | 0.32 | 923 | - | | 18.2; 17.3 |
| Ca | 0.394 | 1115 | - | | 22.41; 22.41 |
| Ba | 0.443 | 1000 | - | | 25.2; 23.9 |
| Ti | 0.293 | 1941 | - | | 23.6; 23.6 |
| Zr | 0.319 | 2182 | - | | 29.3; 29.3 |
| V | 0.2676 | 2183 | - | | 24.6; 24.6 |
| Ta | 0.2914 | 2190 | - | | 26.8;26.8 |
| Fe | 0.252 | 1811 | 9765 | 18.2 | 19.1 |
| Co | 0.2552 | 1728 | - | | 19.3 |
| Ni | 0.2488 | 1728 | 7042 | 17.5 | 18.8 |
| Cu | 0.2555 | 1358 | 890 | 14.9 | 16.1 |
| Zn | 0.2758 | 693 | 1034 | 17.2 | 18.5; 16.5 |
| Pd | 0.2746 | 1825 | 3750 | 19.9 | 20.8 |
| Ag | 0.2884 | 1235 | - | | 17.3 |
| Pt | 0.277 | 1828 | - | | 21.0 |
| Au | 0.288 | 1337 | - | | 17.3 |
| Al | 0.286 | 993 | - | | 16.3;15.4 |
| C | 0.154 | 3800 | - | | 20.5 (m = 2.56) |
| Si | 0.2632 | 1683 | 4 | 9.1 (m = 4.88, f = 0.1) | 10.6 |
| Ge | 0.2732 | 1211 | - | (m = 4.88, f = 0.1) | 11.5 |
| Sn | 0.384 | 505.1 | - | | 47.2; 40.6 |
| Pb | 0.3492 | 600.6 | - | | 28.0; 24.9 |
| Bi | 0.34 | 544.4 | - | | 36.0; 29.2 |
| $Ni_{80}P_{20}$ | 0.2429 | 1184 | 7063 | 7.9 (m = 4) | 8.9 |
| $NiZr_2$ | 0.4681 | 1393 | 7093 | 17.0 (m = 4) | 19.8 |
| $TiO_2$ | 0.3860 | 2098 | 7432 | 22.5 (m = 4) | 23.1(f = 0.01) |

Figure 8 compares the reproduced IHPR with typical measurements for (a) Cu,[164,165,166] (b) Fe,[164] (c) Ni,[75,167] (d) NiP,[168] and NiZr,[76] (e) $TiO_2$,[169] and (f) Si.[7] The straight lines are the traditional HPR and its slope is obtained by adjusting the *f* values in eq (18a). The solid lines are the current BOLS predictions without using any other adjustable parameters. The dashed lines only consider the contribution from the activation energy with $\varphi(d, m, T) = 1$. The scattered data are experimental results reported by different researchers. All the panels were fitted at T = 300 K with f = 0.5 otherwise as indicated. The BOLS predictions match reasonably well to all measurements. The perfect match of NiP alloy and $TiO_2$ may adequately evidence the current BOLS approach that is close to the true situations of IHPR. As can be seen from Table 5, changing the f values from 0.5 to 0.663 has no effect on the critical size for materials with $T_m(0) > 1000$ K, and therefore, for the examined samples, there is no difference using f = 0.5 or 0.663. The small f values for $TiO_2$(0.01) and Si (0.1) may be dominated by the bond nature alteration that raises the $T_m(x_j, m)$.

Figure 8 (link)
Comparison of the calculated (solid line) with the measured IHPR (scattered data) of (a) Cu,[164,165,166] (b) Fe, [164] (c) Ni 75,[167] (d) NiP[168] and NiZr[76] (e) $TiO_2$[169] and (f) Si.[7] The solid curves represent the complete BOLS form of $[1+ f \times x \times \exp(T_m(x)/T)]\varphi(d,m,T)$. The straight lines are traditional HPR representing $1+ f \times x \times \exp(T_m(0)/T)$ and its intercept at the y-axis corresponds to the normalized hardness of the bulk counterparts. The dashed curves lines are the modified HPR with $[1+ f \times x \times \exp(T_m(x)/T)]$. The slope f = 0.5 was optimised otherwise as indicated for all the samples.

Strikingly, a single data point of measurement is sufficient to calibrate the IHPR of a specific solid. Applying the relative hardness of the defect-free 40 nm silicon nanosphere[7] to the IHPR equation results in the maximal hardness of Si nanosolid at room temperature being 5 times the bulk value and the IHPR critical size being 9 ~ 10 nm. The calibration using a single data point should be one of the advantages of the current approach in revealing and calibrating the IHPR of a solid. Applying m = 4.88 for Si to eq (18a) gives immediately $c_1$ = (15 ± 1) % and the corresponding $z_1 \approx 3.60 \pm 0.25$. The $z_1$ for the spherical Si is slightly lower than $z_1 = 4$ for a flat surface and the $c_1$ is slightly higher than that of the TiCrN flat surface (12%) because of the curvature of the sphere.

### 3.6.3 Factors dominating the critical size

Figure 9 compares the size dependence of the ductility (broken lines) and IHPR-2 (solid lines) of (a) Pb (m = 1, $T_m$ = 600.6 K, f = 0.663) and (b) Au (m = 1, $T_m$ = 1337.33 K, f = 0.663) operated at 300K. P(x)/P(0) = 1 corresponds to the critical temperature for solid-semisolid transition; P(x)/P(0) = 0 corresponds to the temperature for semisolid-liquid transition. The normalized compressibility drops first and then bends up at the IHPR critical point until the bulk value at $T_C(x)$ and then goes up to infinity at $T_m(x)$. This trend agrees with that discussed for metallic MC. Results show that Pb nanosolid of x = 0.34 (D = 6 nm) size becomes semisolid and of x = 0.52 (2.6 nm) becomes liquid at 300 K; Au nanosolid of x = 0.68 (D = 1.25 nm) becomes semisolid. The gold MC is in semisolid state, which clarifies further the high extensibility of gold MC at the ambient temperature.[88] The curve of IHPR-1 with φ(x, m, T) = 1 could hardly reach the semisolid or the liquid states in both cases.

Figure 9 ([link](link))
Comparison of the IHPR-2 (solid line) transition and critical sizes for solid-to-semisolid and semisolid-liquid transition at 300 K for (a) Pb (m = 1, $T_m$ = 600.6 K, f =0.663) and (b) Au (m = 1, $T_m$ = 1337.33 K, f = 0.663). $P/P_0$ = 1 corresponds to semisolid state with critical temperature $T_C(x)$; $P/P_0$ = 0 to corresponds to semisolid-liquid transition $T_m(x)$. Broken line represents the reduced compressibility that drops first and then bends up until the bulk value at $T_C(x)$ and then goes up to infinity at $T_m(x)$. The gold MC is in semisolid state at 300K, which clarifies further the high extensibility of gold MC at the ambient temperature.[88] IHPR-1 could not reach the semisolid or the liquid states.

Figure 10 examines the f, m and $T/T_m(0)$ dependence of the $x_C$ values. The critical size varies significantly with the temperature of operation ($T/T_m(0)$ ratio) and the bond nature (m). In the range of f = 0.1 and 1.0, $x_C$ depends less on the f values if the $T/T_m(0)$ ratio is smaller than 0.2.

Figure 10 ([link](link))
Dependence of $x_C$ (m, f, $T/T_m(0)$) on (a) extrinsic factor f and (b) bond nature indicator of m.

### 3.6.4 Size effect on $T_C(x)$ and $T_m(x)$

From the $T/T_m$ and size dependence of strength and ductility in Figure 11a, we may note the following trends in general: (i) For a given material of a given size, the normalized mechanical strength drops from infinity to zero when T approaches $T_m(x, m)$, associated with increase of ductility that becomes singular at the $T_m(x)$ that drops by 5% and 15%, respectively, for $x_C = 0.15$ and 0.21. (ii) Both P and β reach their bulk values (or transits into semisolid state) at $T_C = 0.75 \, T_m$ and $0.65 \, T_m$ for $x_C = 0.15$ and 0.21, respectively. When $T > T_C$, P drops and β rises in an exponential way.

A comparison of the size dependence of the normalized atomic distance $d(x_j)/d(0)$, melting point, $T_m(x_j, m =1)/T_m(0)$, and $T_C(x_j, f, m =1)/T_m(0)$ for solid-to-semisolid transitions shown in Figure 11b indicates that the $T_C(x_j, f)$ drops more rapidly with size than the $T_m(x_j, m)$. The bulk $T_C(0)$ value is about 20% lower than the bulk $T_m$. It is interesting to note that the $T_m(x_j)/T_m(0)$-BOLS curve overlaps the curve of $T_m(x_j)/T_m(0)$-Born, indicating the consistency in the respective physical mechanisms of melting. The former represents that the melting point is governed by the atomic cohesive energy ($T_m \propto z_i E_i$) and the latter representing eq (20) means that the mechanical strength, or the shear modulus, approaches zero at the $T_m$.[90] The two curves also agree with the model of Jiang et al. derived from Lindermann's criterion of atomic vibration and its derivatives of surface lattice/phonon instability.[10,170,171]

Figure 11(link)
(a) The $T/T_m$ ratio dependent strength and ductility of size $x_C = 0.15$ and 0.21, and (b) comparison of size dependence of critical temperature for solid-semisolid transition, $T_C(x)/T_m(0)$, bond length contraction, $d(x)/d(0)$, and semisolid -liquid transition of metallic nanosolid (m = 1). $T_m(x)/T_m(0)$- BOLS curve overlaps the $T_m(x)/T_m(0)$- Born.

For a Cu nanosolid with $K_j = 10$ (5 nm in diameter), for example, the bond contracts by a mean value of 5%, associated with a 25% drop of $T_m$ and a 50% drop of $T_C$ with respect to the bulk $T_m(0)$ value of 1358 K. The 5 nm-sized nanosolid is in a semisolid state at 680 K according to Figure 11b. The self-heating in operation because of bond breaking should raise the actual temperature of the specimen. Hence, the joint effect of size-induced $T_C$ suppression and the bond breaking induced heating may provide an additional mechanism for the high ductility of Cu nanowires at the ambient temperatures.[148]

The predicted m, f, and $T/T_m(0)$ dependence of $x_C$ and $T_C(x)$ and the trends of mechanical strength and compressibility/extensibility coincide exceedingly well with the cases as reported by Eskin, Suyitno, and Katgerman[107] on the grain size dependence of the tensile elongation (extensibility) of an Al-Cu alloys in the semisolid state. The ductility increases exponentially with temperature until infinity at $T_m$ that drops with solid size. On the other hand, the ductility increases generally with grain refinement. This is also the frequently observed cases such as alumina[172] and PbS[173] in nanometer range at room temperature.[174,175] The compressibility of alumina and PbS solid increases, whereas the Young's modulus decreases as the solid size is decreased. The predicted trends also agree with experimental observations[176] of the temperature dependence of the yield stress of Mg nanosolids of a given size showing that the yield strength drops as the operating temperature is elevated.[176] An atomic-scale simulation[177] also suggests that the material becomes softer in both the plastic and elastic regimes as the operating temperature is raised. When measuring at 200 °C, the strength of 300-nm-sized Cu nanograins is lowered by 15% and the ductility increases substantially.[178] Temperature dependence of the tensile properties of ultrafine-grained ordered $FeCo_2V$ samples with grain sizes of 100, 150, and 290 nm revealed that extremely high yield strengths (up to 2.1 GPa) present at room temperature with appreciable ductility of 3 -13%. The measured strengths declined gradually as the testing temperature was increased to 400 °C, while ductility was generally enhanced, up to 22%.[179] At T > 0.7 $T_m$, the mechanical strength decreases rapidly with increasing temperature and with decreasing strain rate.[1] The measured temperature dependence of the Si(111) and Si(100) biaxial Young's modulus shows that the Young's modulus drops linearly when the T is increased.[180,181,182] A large volume of database, as sampled in Figure 12,[107, 183] shows that the tensile strength of alloys drops from the bulk values to zero when the temperature is raised to the melting point, as predicted by Born[90] and the current BOLS correlation mechanism (Figure 11a). Superplasticity that is an excess strain of $10^3$ without any substantial necking region when loaded in tension is generally observed in materials with grain size less than 10 mm in the temperature range 0.5 – 0.6 $T_m$.[184] Large grain boundary area and high self-diffusivity, superplasticity is achievable at lower temperatures and/or higher strain rates for some nanocrystalline materials.

> Figure 12 ([link](link))
> Temperature and Mg composition dependence of the ultimate tensile strength of Al–Mg alloys. Curve labelled 1 – 10 show the trend due to the increase of Mg content from 0.6 to 10%.[183]

IV      Summary

An atomistic and uniscale solution to the mechanical strength and thermal stability for an atomic chain, a surface, and a solid over the whole range of sizes has been developed on the basis of the BOLS correlation with the following facts as physical constraints: (i) mechanical enhancement happens at the site surrounding a defect or at a surface because of the CN imperfection-enhanced bond strength gain; (ii) the molten phase is extremely soft and highly compressible; (iii) atomic dislocation at the grain boundary requires activation energy that is proportional to the melting point; (iv) activation energy for atomic dislocation drops because of the bond order loss. Matching predictions to observations reveals the following:

(i) Without external stimuli, the metallic bond in an MC contracts by ~30%, associated with ~40% magnitude rise of the bond energy. A metallic MC melts at 1/4.185 fold the bulk $T_{m,b}$.

(ii) The strain limit of a bond in a metallic MC under tension varies in-apparently with mechanical stress or strain rate but apparently with temperature separation between $T_m$ and T.

(iii) The large range of deviation in Au-Au distance in measurement originates from thermal and mechanical fluctuations rather than insertion of light atoms.

(iv) It is possible to make an MC of any metals if the operating temperature is carefully controlled in the range of 20 K lower than the $T_{m,b}/4.185$.

(v) The developed approach provides an effective way of determining the bulk 0 K extensibility, $\beta_0$, the effective specific heat $\eta_{1i}$ per coordinate and the energy ($2\eta_{2i}$) required for evaporating an atom from the molten MC, provided with correct $\eta_{1b}$ value.

(vi) At temperature far below the surface $T_m$, a nansolid should be fragile with even lower extensibility, whereas at T approaches the surface $T_m$ or higher, the nanosolid should be ductile.

(vii) For a nanograined nanowire, the bond unfolding and sliding dislocations of the lower-coordinated atoms at grain boundaries are suggested to dominate the super plasticity, as the bond strain limit at temperature close to the melting point is lower than 100%. Understanding should provide an additional mechanism for the high ductility of a metallic nanosolid as the critical temperature for the solid-to-semisolid transition is much lower than the bulk melting point and self-heating should raise the real temperature of the small samples.

(viii) The C-C bond in the SWCNT contracts by ~18.5% with energy rises by ~68%. The effective thickness of the C-C bond is ~0.142 nm, which is the diameter of an isolated C atom. The melting point of the tube-wall is slightly (~ 12 K) higher than that of the open edge of the tube-end. The reported values of tip-end $T_m$ and the product of Yt essentially represent the true situations of a SWCNT in which the Young's modulus is 2.5 times greater and the $T_m$ is 0.42 times greater than that of bulk carbon. Predictions of the wall thickness dependence agree well with the insofar-observed trends in $T_m$ suppression and Y enhancement of the nanobeams.

(ix) A surface is harder at temperatures far below $T_m$ but the surface melts more easily compared to the bulk interior. The temperature separation ($T_m - T$) cannot be neglected in dealing with the mechanical strength and ductility of a nanosolid.

(x) The IHPR originates from the competition between the bond order loss and the associated bond strength gain of atoms in the surface skins. As the solid size is decreased, a transition from dominance of bond strength gain to dominance of bond order loss occurs at the IHPR critical size because of the increased portion of lower coordinated atoms. During the transition, both bond order loss and bond strength gain contribute competitively.

(xi) The IHPR critical size is universally predictable, which can be calibrated with a few measured data points for a specific system. The critical size is dominated intrinsically by bond nature and $T/T_m$ ratio and extrinsically by experimental conditions or other factors such as size distribution and impurities that are represented by the factors f.

(xii) The IHPR at larger solid size converges to the normal HPR that holds its conventional meaning of the accumulation of atomic dislocations that resists further atomic displacements in plastic deformation. The slope in the traditional HPR is proportional to $\exp(T_m/T)$, which represents the relationship between the hardness and the activation energy for atomic dislocations. The x in the conventional HPR should represent the accumulation of atomic dislocations that resists further dislocations, and the constant f may describe the contribution from extrinsic factors such as the shapes of tips, loading scales, strain rates and defect intensity.

Findings provide a consistent insight into the CN-imperfection-enhanced binding intensity, mechanical strength, thermal stability, and the compressibility or extensibility of a MC, metallic nanowires, carbon nanotubes, and the IHPR of a nanosolid. Consistency between predictions and observations, together with progress made by the practitioner and co-workers, further evidence the tremendous impact of atomic CN imperfection and the validity and essentiality of the BOLS correlation mechanism and its derivatives on the mechanical strength of the discussed low dimensional systems and a solid over the whole range of sizes.

It is important to note that, in the nanoindentation test, errors may arise because of the shapes and sizes of the tips, such as in the cases described in ref. 2. In practice, the stress-strain profiles of a nanosolid are not symmetrical when comparing the situation under tension to the situation under compression,[185] and the flow stress is dependent on strain rate, loading mode and time, and materials compactness, as well as particle size distribution. The hardness of single crystal moissanite (6H-SiC) obtained by ten-second loading parallel to the crystallographic *c* axis varies with the loading magnitude. If the load is 0.5 N, the derived hardness is 26 GPa; loading at 29 and 50 N, the corresponding hardness are 22.5 and 22 GPa, respectively.[136] The measured hardness is very complicated, and the hardness of a material is hard to be certain accurately. Fortunately, the effect of tip shape and loading mode never affects the origin and the peaking at the surface in detection. By taking the relative change of the measured quantity into account, artifacts because of the measuring techniques can be minimized in the present approach, seeking for the change relative to the bulk counterpart measured under the same conditions. However, the extrinsic factors could be modelled by changing the prefacer f in the IHPR modelling practice. The relative change of intensity, the peak position, and the trend of change could reflect the intrinsic characteristics in physics.

Thermal energy released from bond breaking and bond unfolding should add energy to the system by raising the actual temperature of the system. The bombardment of electron beams in the TEM will lower the $T_m$ by hundreds of degrees depending on the beam energy.[60] Furthermore, the fluctuation in grain size distribution also affects the actual melting point of the individual atoms at boundaries of grains of different sizes. These factors may explain why the NW becomes thinner and thinner when it is being stretched at a broad temperature range.

One may note that the Young's modulus is defined for elastic deformation, while its inverse, or the extensibility or compressibility, covers both the elastic and plastic types of deformation. However, either the elastic or the plastic process is related to the process of bond distortion, including bond unfolding, stretching, or breaking, that consumes energy (obtained by integrating the stress with respect to strain) being a certain portion of the entire binding energy. No matter how complicated the actual process of bond deformation (with linear or nonlinear response) or recovery (reversible or irreversible) is, a specific process consumes a fixed portion of bond energy, and the exact portion for the specific process does not come into play in the solution seeking for relative change. Therefore, the BOLS and its derivatives are valid for any substances of any scales of size and in any phases even liquid and gaseous. As the Young's modulus and mechanical strength are quantities of intensity, they are volumeless. Therefore, it might not be appropriate to think about the stress of a single atom, instead, the stress at a specific atomic site.

Nevertheless, one could not expect to cover fluctuations due to mechanical (strain rate, stress direction, loading mode and time, etc.), thermal (self-heating during process), crystal structure orientation, impurity density, or grain-size distribution effects in a theoretical model, as these fluctuations are extrinsic and hardly controllable. We should focus more on the nature and trend of the unusual behaviour in mechanics, as accurate detection of the absolute values remains problematic.[136]

As the BOLS correlation deals with only the effect of bond order loss, none of the particularities of the elements, crystal or phase structures or the form of pair potential is involved. What we need to consider are the nature of the bond and the equilibrium atomic distance with and without external stimulus. This could be one of the advantages of the reported approach. Therefore, no multiscale approaches are necessary in the current "bottom up" exercise.

Further extension of the BOLS premise should open up a new way of thinking about nanostructures that are dominated by the effect of bond order loss. Miniaturization of dimensionality not only allows us to tune the physical properties of a solid but also provides us with opportunity to elucidate information that is beyond the scope of traditional approach. Gaining information includes the energy levels of an isolated atom, the vibration frequency of an isolated dimer bond, and the specific heat per bond, as well as discrimination the contribution of bond order loss from the effect of chemical passivation, may form the subject of nanometrology.

Transport dynamics of phonons, electrons, and photons in the nanometre regime is also a high challenge. Densification of mass, charge, and electrons in the surface edge or grain boundaries is expected to lead to new phenomena that have high potential for practical application in design of new functional materials and devices.

Acknowledgment

Profitable communications with Professors Stan Viperek, Chunli Bai, John S. Colligon, S.R.P. Silva, David S. Y. Tong, F.M. Ashby, Philip J. Jennings, Andris Stelbovics, A. Bhalla, and En-Yong Jiang are all gratefully acknowledged.

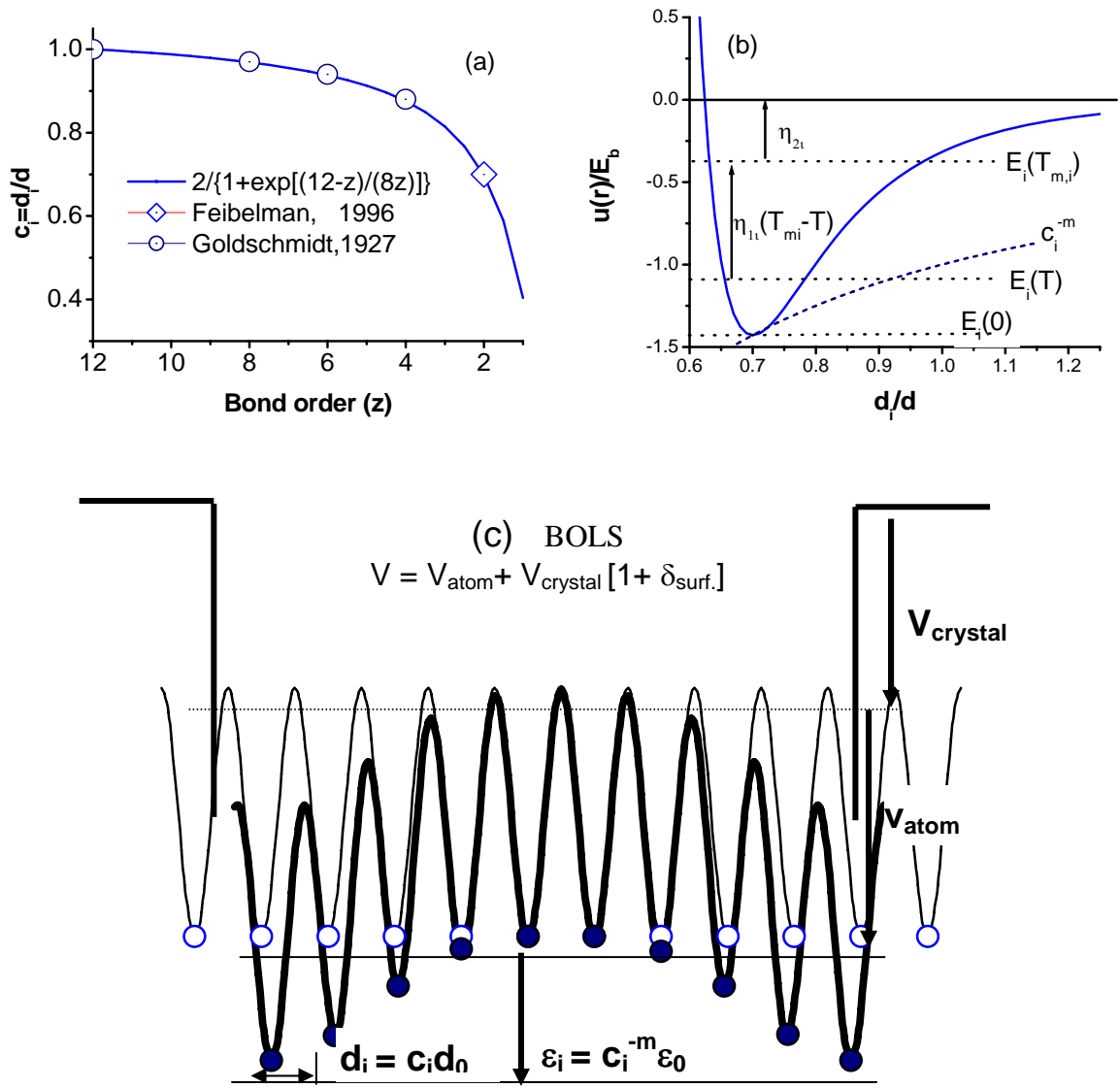

Fig-01

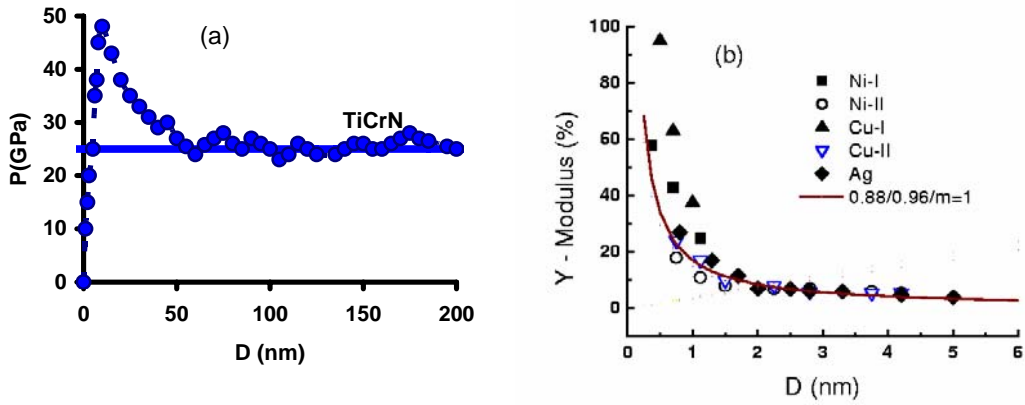

Fig-02

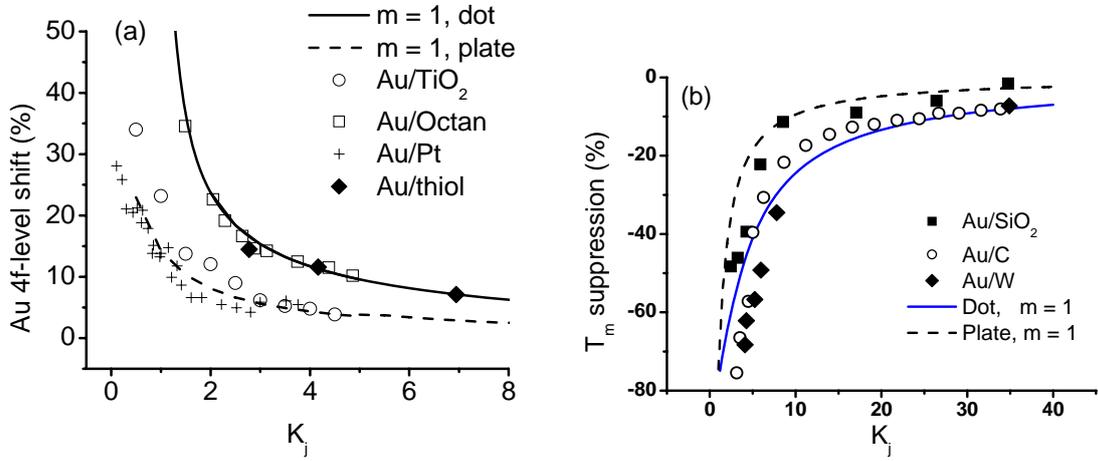

Fig-03

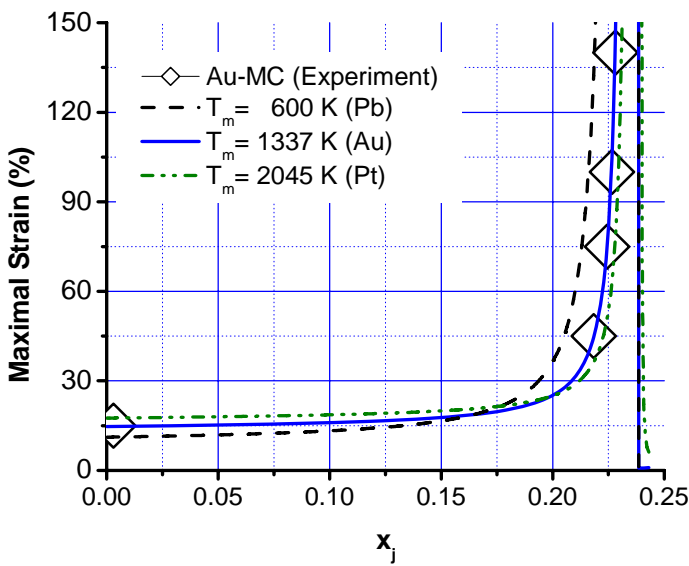

Fig-04

Fig-05 (counter plot is ignored)

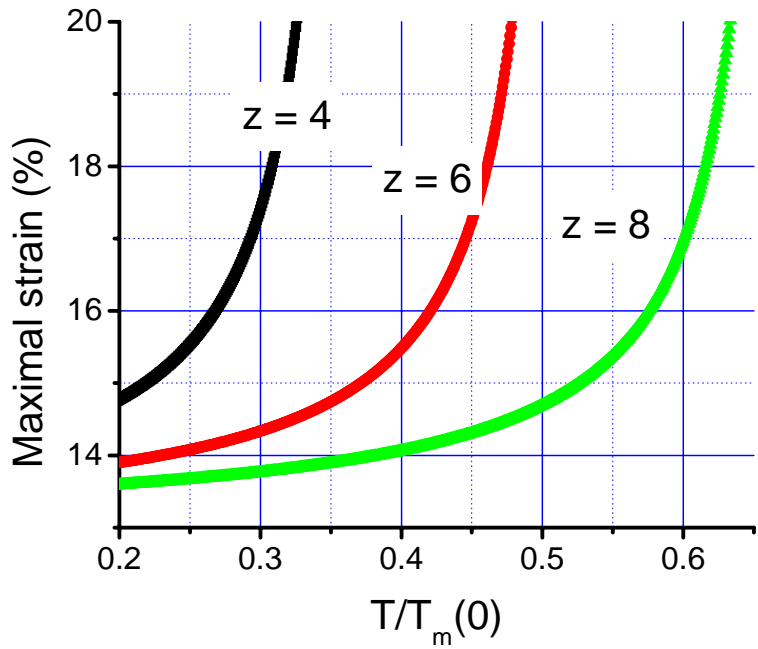

Fig-06

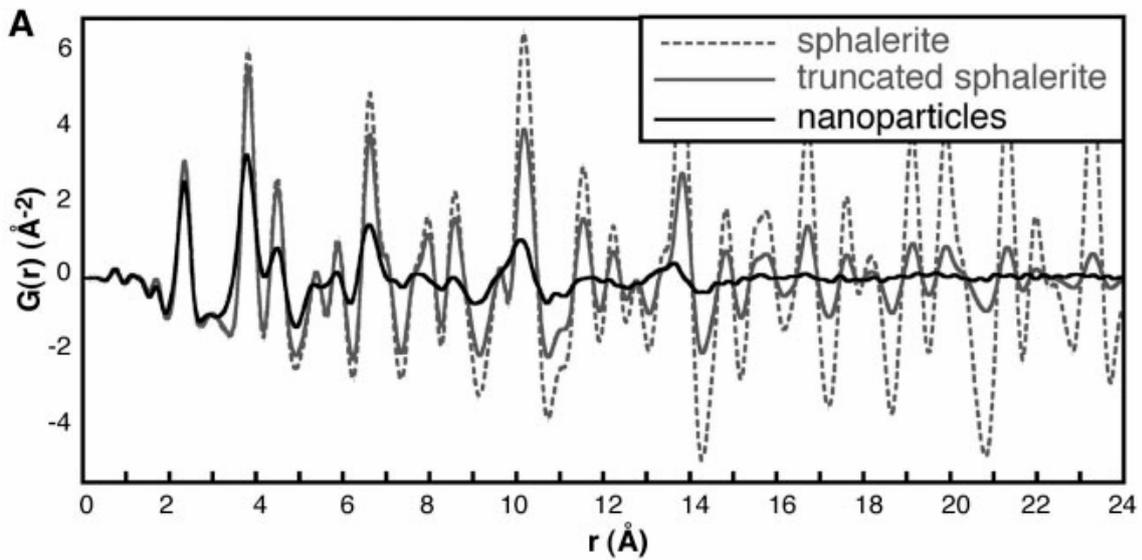

Fig-07

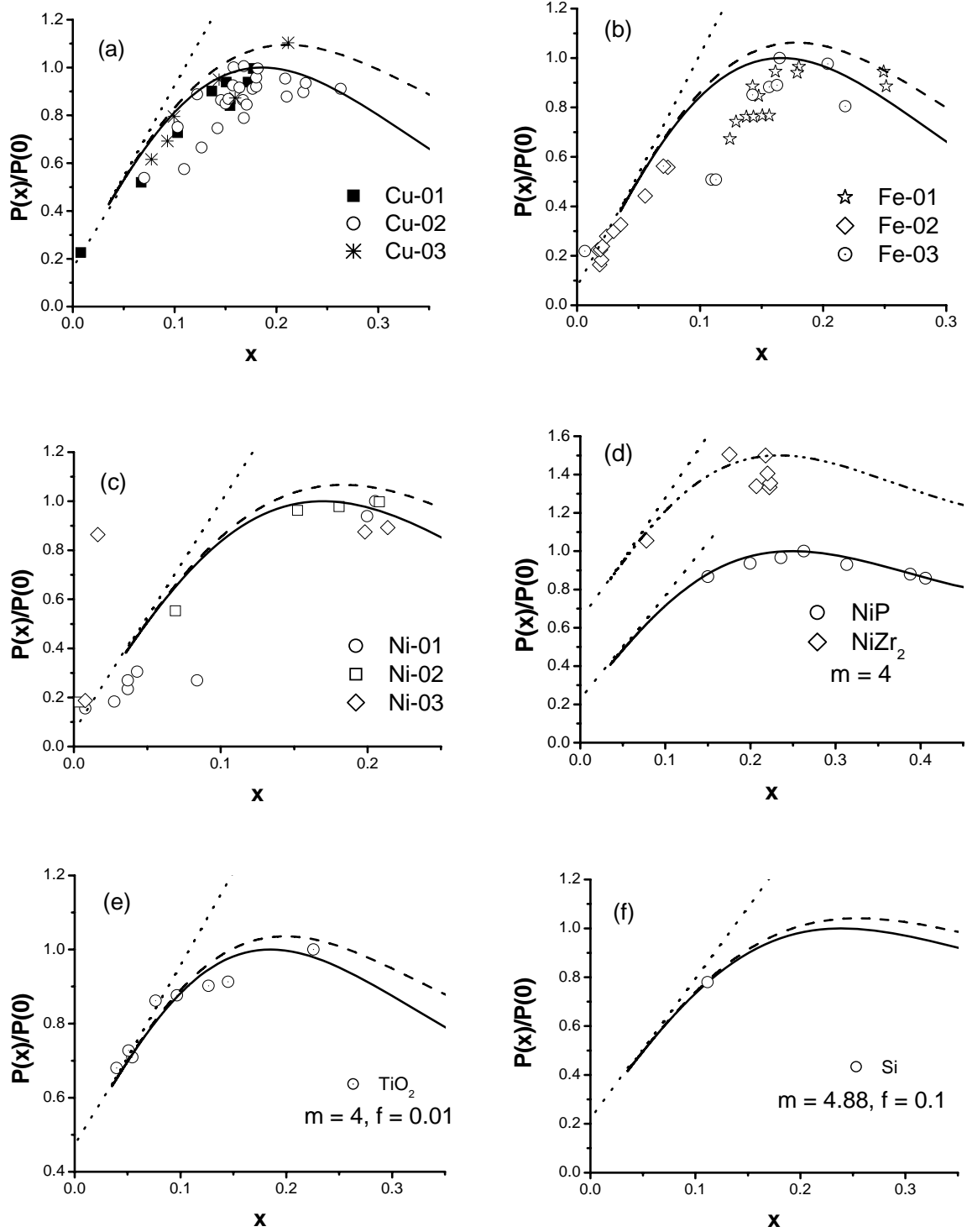

Fig-08

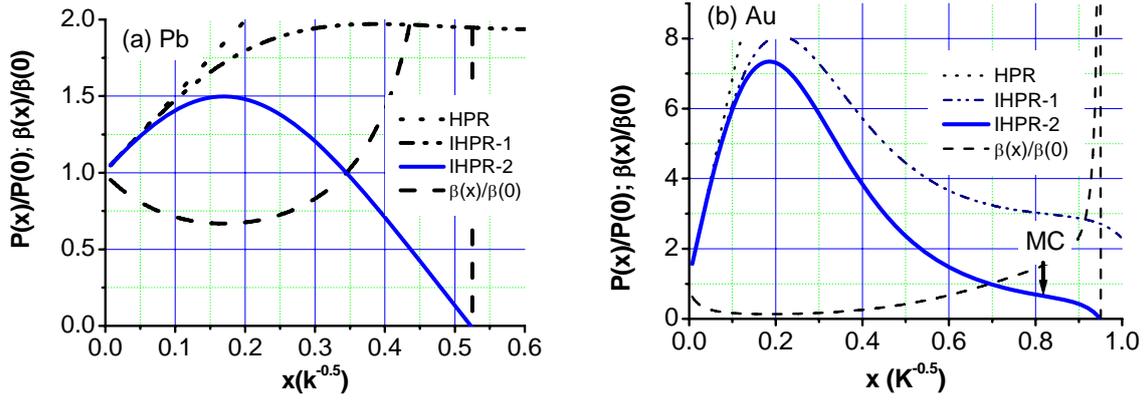

Fig-09

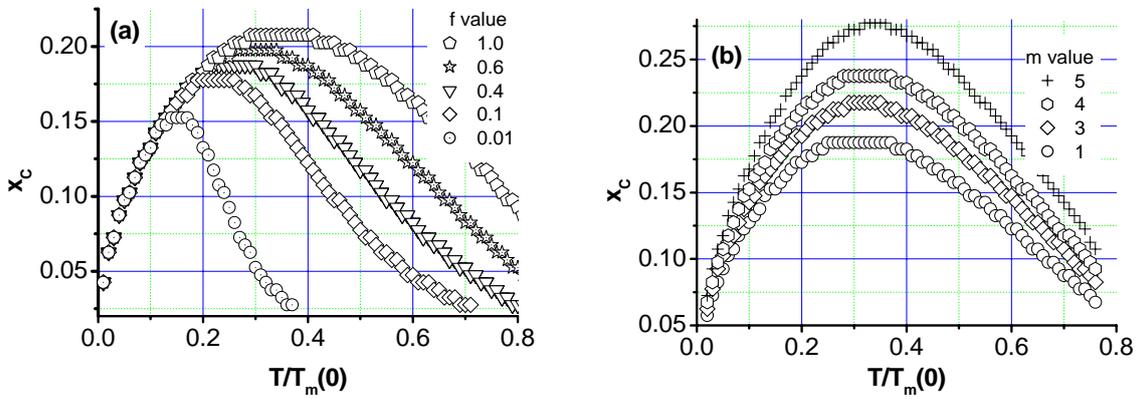

Fig-10

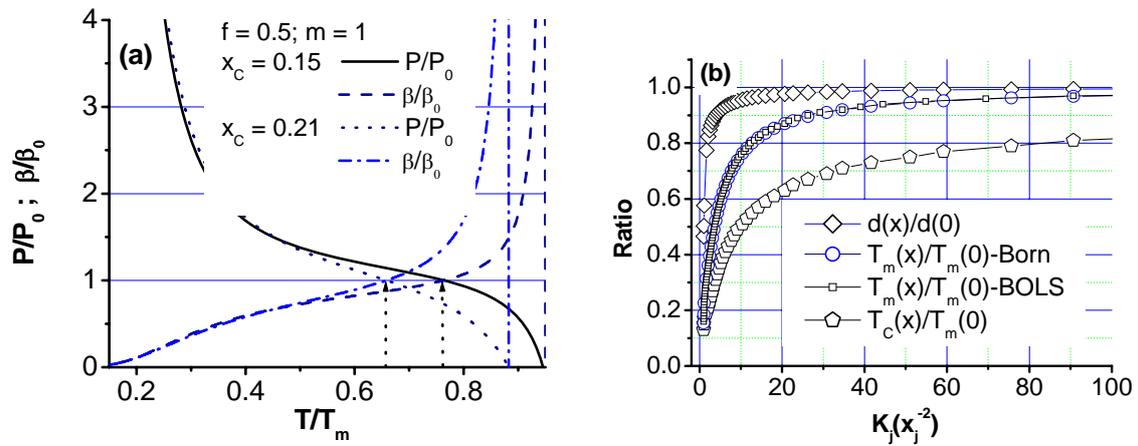

Fig-11

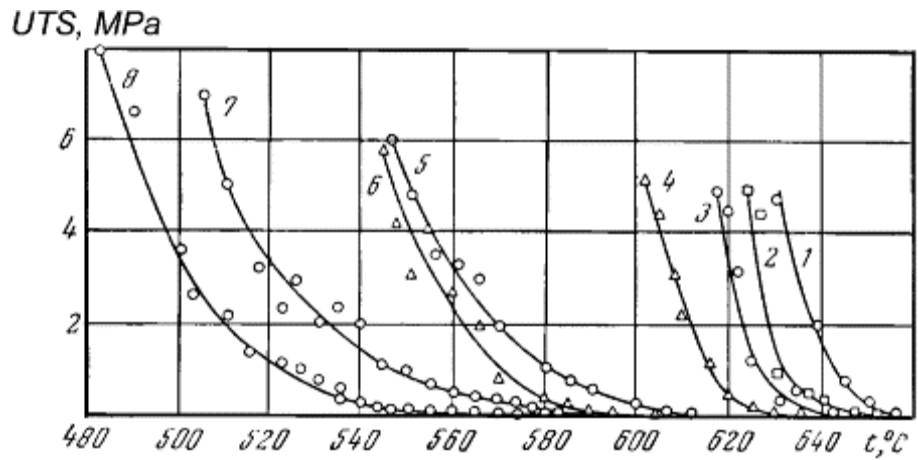

Fig- 12